\documentclass[a4paper,11pt]{article}
\pdfoutput=1 

\usepackage{jcappub}

\usepackage[T1]{fontenc} 

\def\bea{\begin{eqnarray}}
\def\eea{\end{eqnarray}}
\def\be{\begin{equation}}
\def\ee{\end{equation}}
\def\v{\mathbf}
\def\mean#1{\left< #1 \right>}

\usepackage{float}
\usepackage[caption = false]{subfig}

\usepackage{multirow}

\usepackage[dvipsnames]{xcolor}

\newcommand{\hMpc}{h^{-1}\rm{Mpc}}

\begin{document}

\title{Baryon Acoustic Oscillations reconstruction with pixels}

\author[a,d]{Andrej Obuljen,}
\author[b,c,d]{Francisco Villaescusa-Navarro,}
\author[e,f]{Emanuele Castorina,}
\author[c,d]{Matteo Viel}

\affiliation[a]{SISSA- International School for Advanced Studies, Via Bonomea 265, 34136 Trieste, Italy}
\affiliation[b]{Center for Computational Astrophysics, 160 5th Ave, New York, NY, 10010, USA}
\affiliation[c]{INAF, Osservatorio Astronomico di Trieste, via Tiepolo 11, I-34131 Trieste, Italy}
\affiliation[d]{INFN -- National Institute for Nuclear Physics, Via Valerio 2, I-34127 Trieste, Italy}
\affiliation[e]{Berkeley Center for Cosmological Physics, University of California, Berkeley, CA 94720, USA}
\affiliation[f]{Lawrence Berkeley National Laboratory, 1 Cyclotron Road, Berkeley, CA 93720, USA}

\emailAdd{aobuljen@sissa.it}
\emailAdd{fvillaescusa@simonsfoundation.org}
\emailAdd{ecastorina@berkeley.edu}
\emailAdd{viel@oats.inaf.it}

\abstract{Gravitational non-linear evolution induces a shift in the position of the baryon acoustic oscillations (BAO) peak together with a damping and broadening of its shape that bias and degrades the accuracy with which the position of the peak can be determined. BAO reconstruction is a technique developed to undo part of the effect of non-linearities. We present and analyse a reconstruction method that consists of displacing pixels instead of galaxies and whose implementation is easier than the standard reconstruction method. We show that this method is equivalent to the standard reconstruction technique in the limit where the number of pixels becomes very large. This method is particularly useful in surveys where individual galaxies are not resolved, as in 21cm intensity mapping observations. We validate this method by reconstructing mock pixelated maps, that we build from the distribution of matter and halos in real- and redshift-space, from a large set of numerical simulations. We find that this method is able to decrease the uncertainty in the BAO peak position by 30-50\% over the typical angular resolution scales of 21 cm intensity mapping experiments.}


\maketitle

\section{Introduction}
\label{sec:introduction}

The standard model of cosmology describes the energy content of the Universe as the sum of different contributions: $\sim 5\%$ of normal (baryonic) matter, $\sim25\%$ of cold dark matter (CDM) and $\sim70\%$ of the mysterious dark energy. N-body simulations have shown that gravitational non-linear evolution tends to organize matter in dense halos, connected among themselves through filaments surrounded by huge voids in the so-called \textit{cosmic web}. The statistical properties of the distribution of matter in the cosmic web contain a huge amount of information such as the fraction of the different energy components, the properties of the density field after inflation or the geometry of the Universe. 

Unfortunately, the cosmic web can not be observed directly, since it is mainly made up of CDM, that does not emit any light or interact with baryonic matter besides gravity. However, there are several cosmological observables that can be used to trace the spatial distribution of the underlying matter density field: the spatial distribution of galaxies, the magnification and shearing of background galaxies by weak gravitational lensing, the spatial distribution of neutral hydrogen as traced by the Ly$\alpha$-forest or via 21cm intensity mapping observations. 

The idea of using tracers of the underlying matter density field, as galaxies or cosmic neutral hydrogen, to constrain the values of the cosmological parameters is based on the reasonable assumption that baryonic overdensities should trace dark matter overdensities on large scales. In other words, in places where there are large matter densities, as in dark matter halos, it is expected that there will also be large baryonic overdensities that will give rise to the formation of galaxies which could also contain neutral hydrogen. 

Under the above assumption one will expect that the clustering patterns of galaxies and matter will be the same on large scales, modulo an overall normalization: the \textit{galaxy bias}. Therefore, measurements of the shape of the galaxy power spectrum or correlation function can be used to constrain the value of the cosmological parameters. 

A different way to constrain the value of some of the cosmological parameters is through \textit{standard rulers}: objects or structures with an intrinsic size that do not change with time. The most prominent example is the baryon acoustic oscillations (BAOs) whose origin reside in the fact that perturbations in the early Universe produced sound waves in the baryon-photon plasma that propagated until recombination, after which photons were not coupled to baryons and they free-streamed until today.  On the other hand, the distribution of baryons stalled at recombination. This phenomenon leaves its signature on the spatial distribution of matter at late times \cite{Eisenstein_Hu_1998}, where the probability of finding two galaxies separated by a particular scale, \textit{the sound horizon}, that is very well constrained by CMB observations, is enhanced. By measuring that scale in galaxy surveys it is possible to measure the value of the Hubble function, $H(z)$, and the angular diameter distance, $D_A(z)$, at redshift $z$.

One of the advantages of BAOs is its robustness against systematic effects, that can impact more strongly other cosmological observables such as those making use of the shape of the galaxy clustering pattern. BAOs produce a peak in the correlation function at $r\sim100~h^{-1}{\rm Mpc}$ while in the power spectrum it produces a set of wiggles at $k\gtrsim0.01~h{\rm Mpc}^{-1}$. Unfortunately, non-linear gravitational evolution produces a damping, broadening \cite{Meiksin_1999, Eisenstein_2007a, Matsubara_2008} and also induces a shift \cite{Crocce_2008, Padmanabhan_2009b,McCullagh_2012} in these features that 1) makes more difficult the task of measuring the sound horizon and 2) could bias the inferred cosmological quantities.  Non-linearities at BAO scales are also sensitive to 
the presence of massive neutrinos as shown in \cite{peloso15}.

A technique to overcome, or at least to mitigate this problem, has been recently developed and it is called \textit{reconstruction} \cite{Eisenstein_2007b, Seo_2008,Padmanabhan_2009,Noh_2009, Tassev_2012,White_2015}. The underlying idea is that non-linear gravitational clustering on BAO scales can be accurately modeled by perturbation theory, and in particular, with Lagrangian perturbation theory: the Zel'dovich approximation \cite{Carlson_2013, White_2014} in its simplest version. Nowadays, BAOs are routinely used for quantitative cosmological  investigations (e.g. \cite{auborg15, Alam_2016,Zhao_2016,Beutler_2016}).

The reconstruction technique can be applied to galaxy surveys, where the goal is to try to move back galaxies to their initial positions (or equivalently to move information embedded into higher order correlations back to the two-point function \cite{Schmittfull2015}). However, there are some observables, like 21cm intensity mapping maps where the output of observations does not consist of a catalogue with the positions of galaxies on the sky, but pixelated maps. For this type of observations the standard reconstruction technique can not be applied, although one possibility would be to use Eulerian reconstruction techniques (see \cite{Schmittfull2015}). Recently, a mesh-based method has been proposed \cite{Seo2015} to carry out BAO reconstruction from 21cm interferometry observations in the presence of foreground contaminations and has been tested at the level of the propagator.

In this paper we provide a more complete and detailed study of the mesh-based reconstruction technique that can be applied to both galaxy surveys and 21cm single-dish intensity mapping observations. This method is similar to the standard reconstruction technique, but relies on moving pixels instead of points (galaxies). By using a large set of numerical simulations we create mock pixelated maps from the distribution of both matter and halos in both real- and redshift-space. We then apply our method to those maps and investigate the performance of the method. We also demonstrate that in the case of galaxy surveys, this method is equivalent to standard reconstruction in the limit of very small pixels.

This paper is organized as follows. In section \ref{sec:Method} we investigate the impact of instrumental effects on the amplitude and shape of the BAO peak from observations consisting on pixelated maps, focusing our attention for concreteness on the case of 21cm intensity mapping. In section \ref{sims} we outline the simulations used in this work, while in section \ref{algorithm} we describe the reconstruction algorithm.
Our theoretical model and the methods we use to fit the results from the simulations is described in section \ref{sec:analysis}. We show and discuss the main results of this paper in section \ref{sec:results}. Finally, the main conclusions of this work are presented in section \ref{sec:conclusions}.

\section{Pixelated maps observations and instrumental effects}
\label{sec:Method}

We begin this section by describing briefly one type of cosmological survey that produces as output pixelated maps: single-dish 21cm intensity mapping observations. We then investigate the impact of instrumental effects, that determine the resolution of the pixelated maps, onto the shape and amplitude of the BAO peak as inferred from those maps. The goal of this section is to study the impact of maps resolution on the monopole and quadrupole of the correlation function and incorporate those effects into our theoretical template that we will use to determine the position of the BAO peak from our mock maps.

\subsection{Pixelated observations: intensity mapping}

An example of surveys producing pixelated maps is given by 21cm intensity mapping observations \cite{Bull2015a,pacobao,chime14}. The intensity mapping technique consists in carrying out a low angular resolution survey, where individual galaxies or HI blobs are not resolved, to measure the 21cm radiation from cosmic neutral hydrogen from large patches of the sky. The idea is the same as with galaxy surveys: HI perturbations on large-scales will trace the underlying matter perturbations. There are two types of observations that can be carried out with radio-telescopes: single-dish or interferometry. In this paper we focus our analysis on single-dish autocorrelation observations\footnote{We refer the reader to \cite{Seo2015} for a study in reconstruction with interferometry observations.}, where the resolution of the maps depends on the size of the antennae and where the maximum angular transversal scales that can be probed are not limited by the field-of-view (FoV) of the radio-telescope. However, we stress that low angular resolution is a limiting factor also for interferometry. A detailed description of the pros and cons of the two different techniques can be found in \cite{Bull2015a,pacobao}. A way to perform reconstruction by combining observations from 21cm intensity mapping and galaxy redshift surveys has also been suggested in order to fill in the missing modes lost due to the foregrounds contaminations \cite{Cohn2015}.

In single-dish radio surveys the angular resolution of the 21cm maps is given by $\theta_{\rm FWHM}\cong\lambda/D$, where $\lambda=0.21(1+z)\;$ m  is the wavelength (in meters) of the 21cm radiation and $D$ is the diameter of the antenna. We assume for simplicity that the primary beam of the telescope is well described by a Gaussian, thus the measured temperature on the $\hat{n}$ direction of the sky is 
\be
\tilde{T}_b(\hat{n},\nu)=\int d\vec{s}_\bot T_b(\nu,\vec{s}_\bot)\frac{1}{2\pi \sigma^2}e^{-s_\bot^2/2\sigma^2} \, ;
\ee
in Fourier space the amplitude of the modes will be given by
\be
\tilde{\delta T}_b(k_\|,\vec{k}_\bot)=e^{-k_\bot^2\sigma^2/2}\delta T_b(k_\|,\vec{k}_\bot) \, ,
\ee
where $\tilde{\delta T}_b$ and $\delta T_b$ represent the observed and cosmological modes and the comoving angular smoothing scale, $\sigma$, is given by
\be
\sigma=\frac{\chi(z)\theta_{\rm FWHM}}{2\sqrt{2\log{2}}}
\ee
with $\chi(z)$ being the comoving distance to redshift $z$ and the factor $2\sqrt{2\log{2}}$ is due to the relation between the FWHM and the standard deviation in the Gaussian function. We notice that while in real observations the density of pixels in a map is closely related to the map resolution, in this paper we consider these two quantities separately. The reason, as we will see clearly on section \ref{subsec:algorithm}, is that the number of pixels can be taken arbitrarily high and this has some benefits for reconstruction. We emphasize that the important quantity in our study is the angular resolution of the maps, parametrized by the parameter $\sigma$.

\subsection{Instrumental effects}

\begin{figure}[htbp]
\centering
\subfloat{\includegraphics[scale = 0.43]{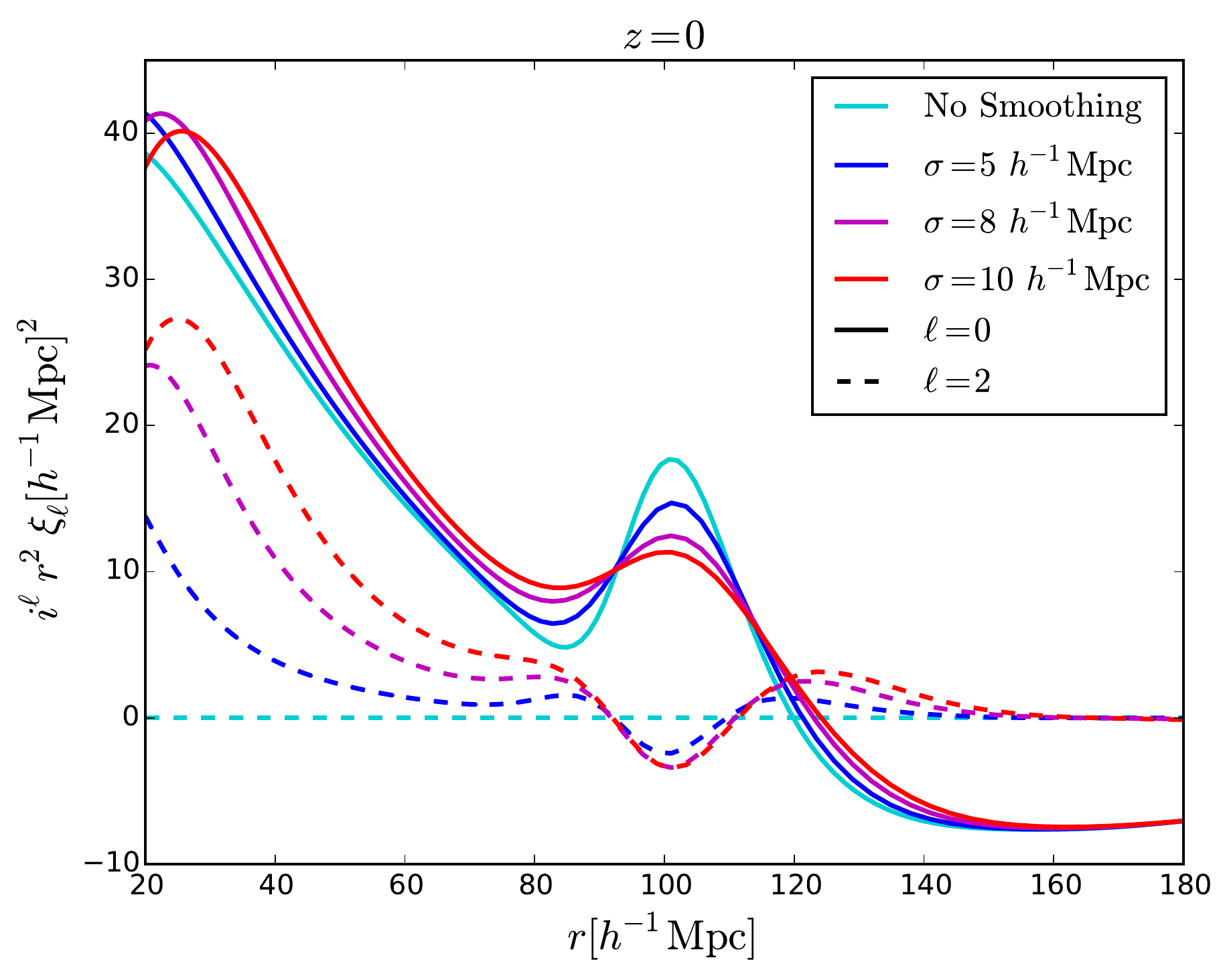}}
\subfloat{\includegraphics[scale = 0.43]{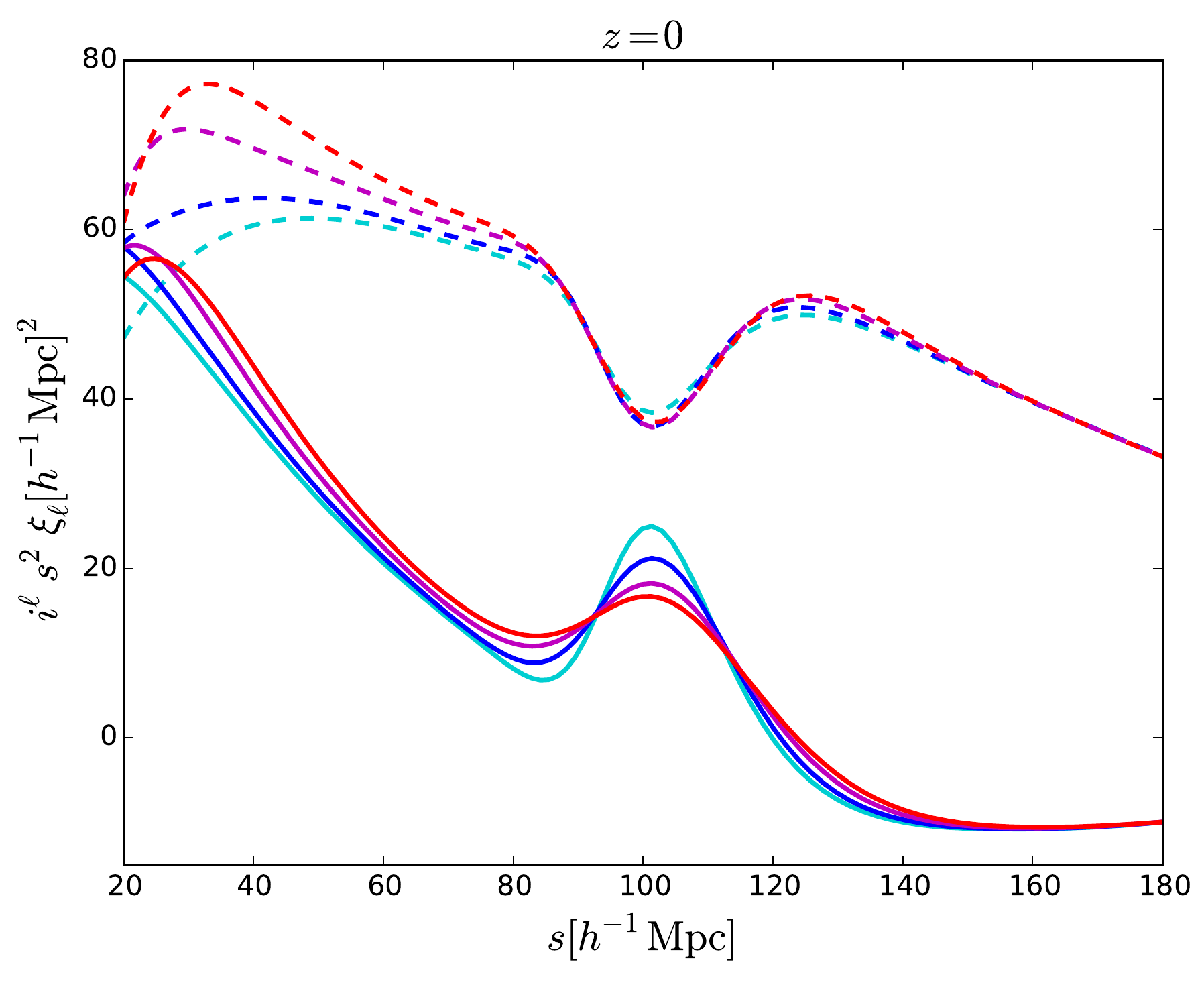}}
\caption{Linear theory prediction for matter monopole (solid) and quadrupole (dashed) of the correlation function for various smoothing scales in real- (left panel) and redshift-space (right panel) at $z=0$.}
\label{fig:Linear_prediction}
\end{figure}

The amplitude and shape of the BAO peak is affected by both non-linear gravitational evolution and instrumental effects. While the goal of this work is to develop a method to undo, at least partially, the effect of non-linearities, the effects induced by the instrument may not be taken out. An example is given by the maps resolution from single-dish 21cm intensity mapping observations, an effect induced by the antenna diameter and that the presence of system noise avoids the possibility of deconvolving the signal to recover to underlying field (see for instance \cite{pacobao}). In this situation, it is important to incorporate the instrumental effects on the BAO peak when building up the theoretical template. We limit our analysis to the impact of resolution on the shape and amplitude of the BAO peak, but we notice that other effects, such as system noise and foregrounds contamination can also affect it (see \cite{pacobao} for a study where these effects were taking into account when detecting the radial BAO). The aim of this subsection is to show the effects induced by the resolution of the pixelated maps on the BAO feature. 

Low angular resolution of the radio telescopes is one of the instrumental effects that must be taken into account when fitting for the position of BAO. Within the framework described above, the 21cm power spectrum in redshift-space from single-dish observations in linear theory is given by:
\be
\tilde{P}_{\rm 21cm}(k,\mu)=b_{\rm 21cm}^2(1+\beta\mu^2)^2e^{-(1-\mu^2)k^2\sigma^2}P_{\rm m}(k),
\ee
where $P_{\rm m}(k)$ is the linear matter power spectrum, $\mu=k_\|/k$ and $b_{\rm 21cm}$ is the bias of the 21cm signal, which is given by $b_{\rm 21cm}=\bar{T}_bb_{\rm HI}$, with $b_{\rm HI}$ being the HI bias and $\bar{T}_b$ is the mean brightness temperature
\be
\bar{T}_b(z)=190\frac{H_0(1+z)^2}{H(z)}\Omega_{\rm HI}(z)h~{\rm mK},
\ee
where $H(z)$ is the value of the Hubble parameter at redshift $z$ and $\Omega_{\rm HI}(z)=\rho_{\rm HI}(z)/\rho_c^0$. We notice that even in real-space ($\beta=0$), the measured 21cm power spectrum is not isotropic, since this symmetry is broken by the angular smoothing in the angular direction. The multipoles of the observed 21cm power spectrum are given by
\be
\tilde{P}_{\rm 21cm,\ell}(k)=\frac{2\ell+1}{2}\int_{-1}^1 L_\ell(\mu)\tilde{P}_{\rm 21cm}(k,\mu)d\mu,
\ee
where $L_\ell(x)$ is the Legendre polinomial of order $\ell$. The multipoles of the observed 21cm correlation function can be written as a function of their power spectrum counterparts
\be
\tilde{\xi}_{\rm 21cm,\ell}(r)=i^\ell\int_0^\infty \frac{k^2dk}{2\pi^2}\tilde{P}_{\rm 21cm,\ell}(k)j_\ell(kr),
\ee
with $j_\ell(x)$ being the spherical Bessel function of order $\ell$. 
In Fig. \ref{fig:Linear_prediction} we plot the monopole and quadrupole of the observed 21cm correlation function in real- and redshift-space at linear order for 21cm maps having different resolutions (characterized by the parameter $\sigma$). For simplicity in Fig. \ref{fig:Linear_prediction} we have taken $b_{\rm 21cm}=1$ mK and $b_{\rm HI}=1$. It can be seen that the shape of the BAO peak gets distorted by the map resolution; the effect is similar to the one induced by non-linearities, i.e. the BAO peak gets damped and broader. This distortion increases with 
 $\sigma$, both in real- and redshift-space. This happens because the smoothing in the transverse direction erases correlations on angular scales smaller than $\sim \sigma$. For angular smoothing scales $\sigma\gtrsim10~h^{-1}$Mpc the BAO feature will be almost completely erased in the monopole of the correlation function, however the BAO peak can still be seen in the radial 21cm power spectrum \citep{pacobao}, although the amount of information embedded there is much smaller. 

As expected in linear theory, the quadrupole in real-space is zero when no angular smoothing is applied. On the other hand, for values of $\sigma$ larger than zero the quadrupole deviates significantly from zero in real-space. The reason is that the angular smoothing breaks the isotropy present in real-space, inducing a non-negligible quadrupole that increases with $\sigma$. In redshift-space the amplitude of the quadrupole on large scales arising from the Kaiser term is much larger that the one induced by the angular smoothing, so the impact of the telescope angular resolution does not modify significantly the shape and amplitude of the quadrupole on scales $r\gtrsim80~h^{-1}$Mpc. On smaller scales the angular smoothing becomes more important, with the amplitude of the quadrupole increasing with $\sigma$.

\section{Simulations}
\label{sims}
\label{sec:sims}

Generating mock 21cm maps is computationally very challenging since large box-size high-resolution hydrodynamic simulations, coupled to radiative transfer calculations, are needed to properly simulate the spatial distribution of neutral hydrogen in the post-reionzation era. A computationally less expensive alternative, although less precise, way consists in populating dark matter halos with neutral hydrogen a-posteriori \cite{Villaescusa_14}. The way dark matter halos are populated with HI can be calibrated using hydrodynamic simulations with small box sizes or by means of analytic models that reproduce the observational data \cite{Villaescusa-Navarro2016, Castorina_16}. The idea of this method is thus to run a standard N-body simulation to obtain a catalogue of dark matter halos which are populated with HI during the post-processing. While N-body simulations are much faster than hydrodynamic simulations, the simulation set this work requires (500 simulations) made this method computationally unfeasible given the computational resources we have access to.

Many different methods have been developed such as {\sc PTHALOS} \cite{PTHALOS}, Augmented Lagrangian Perturbation Theory (ALPT) \cite{Kitaura_2013}, PerturbAtion Theory Catalog generator of Halo and galaxY distributions (PATCHY) \cite{PATCHY}, Comoving Lagrangian Acceleration method (COLA) \cite{Tassev_2013,Tassev_2015,L-PICOLA}, Effective Zel'dovich approximation mocks (EZmocks) \cite{EZmocks}, FastPM \cite{FastPM} and {\sc PINOCCHIO} \cite{Monaco_2002, Taffoni_2002, Monaco_2002b, Monaco_2013} (see \cite{Chuang_2015} for a comparison among the different methods and N-body simulations) that are able to either generate a mock distribution of dark matter halos or to evolve directly the matter distribution in an N-body manner. These methods make use of different approximations that determine, in most cases, the accuracy they can reach when comparing results against N-body simulations. 

The rationale behind the use of the above methods is to generate halo catalogues and simulate the spatial distribution of matter into the fully non-linear regime in a faster way than an N-body simulation. Among the previous methods, we have chosen COLA to run our numerical simulations. COLA is basically a particle-mesh (PM) code and therefore can be considered as an N-body code. The difference with respect to a fully N-body is the number of times steps and the way COLA deals with small scales. Given the high accuracy COLA can reach and the fact that it is computationally much faster than an N-body simulation, we decide to use this code to run our numerical simulations. 

We have run 500 simulations using the publicly available {\sc L-PICOLA} code \cite{L-PICOLA}, a version of the original {\sc COLA} code \cite{Tassev_2013}. In the simulations we follow the evolution of $512^3$ dark matter particles within a box of side $1~h^{-1}{\rm Gpc}$ from $z=9$ to $z=0$ using a grid with a number of cells equal to the number of particles. The values of the cosmological parameters we use for all simulations are: $\Omega_{\rm m}=0.3175$, $\Omega_{\Lambda}=0.6825$, $\Omega_{\rm b}=0.049$, $\Omega_\nu=0.0$, $h=0.6711$, $n_s=0.9624$, $\sigma_8=0.834$. We save snapshots at $z=1$ and $z=0$. The outputs at $z=1$ are obtained using 10 time-steps, while we use 50 time-steps linearly spaced in scale factor $a$ for outputs at $z=0$. The mass resolution of the dark matter particles is $6.56\times10^{11}~h^{-1} \rm{M_\odot}$. We identify dark matter halos using the Friends-of-Friends algorithm \cite{FoF} with a linking length parameter $b=0.2$. Halos containing less than 32 dark matter particles ($M_{\rm halo}\lesssim2\times 10^{13} h^{-1} \rm{M_\odot}$) are discarded from our catalogues.

\subsection{Creating mock maps}
\label{subsec:maps}
Here we explain how we simulate the intrinsic resolution of the 21cm maps in our simulations. From the output of the numerical simulations we build mock pixelated maps using the distribution of matter or halos in both real- and redshift-space. We compute the overdensity field $\delta(\v{x})$ of particles in a simulation on a regular grid using cloud-in-cell (CIC) scheme. We Fourier transform $\delta(\v{x})$ to obtain $\tilde{\delta}(\v{k})$ and we correct for the CIC mass assignment scheme. We then apply a transverse 2D Gaussian filter to the density field with an angular smoothing scale $\sigma$:
\begin{equation}
\tilde{\delta}_{\rm{sm}}(\v{k})=\tilde{\delta}(\v{k})e^{-k^2(1-\mu^2)\sigma^2/2}.
\end{equation}

We varied the angular smoothing scale within a reasonable range of values $\sigma=5,8,10\allowbreak \,\allowbreak \hMpc$, that we use throughout the paper. We call the resulting fields -- matter and halo maps.

We emphasize that the matter and halo maps constructed following the above procedure do not correspond to actual 21cm intensity mapping maps (see for instance \cite{Villaescusa_14}). The goal is to create pixelated maps with different levels of complexity, i.e. incorporating redshift-space distortions, halo bias...etc, in order to investigate the robustness of our method against these processes. 

When analysing the matter maps in redshift-space, in each realisation we measure the monopole and the quadrupole along three different axes of our simulation and use the average monopole and quadrupole.

To create a pixelated map from a galaxy survey, in which the angular resolution effects are negligible, we follow the procedure described above and set the angular smoothing scale $\sigma=0$.

\section{Reconstruction algorithm}
\label{algorithm}
We start this section by explaining how the standard reconstruction method works. We then describe in detail our pixelated BAO reconstruction algorithm together with its practical implementation.

\subsection{Standard reconstruction}
The density field reconstruction method was first presented in \cite{Eisenstein_2007b} and it has proved very successful in both data \cite{Anderson2012,Padmanabhan2012,Anderson2014a} and simulations \cite{Seo_2008,Seo2010,Noh_2009,Mehta2011}. Here we summarise the method briefly to set up notation and outline the general idea. 

A position of a particle in Eulerian coordinates $\v{x}$ after time $t$ can be mapped to the initial Lagrangian position $\v{q}$ using the displacement field $\v{\Psi}(\v{q},t)$:
\be
\v{x}(\v{q},t)=\v{q}+\v{\Psi}(\v{q},t).
\ee

Lagrangian Perturbation Theory (LPT) gives a perturbative solution for this displacement field and the first order solution is the Zel'dovich Approximation (ZA) \cite{Zeldovich}. In ZA we can express the overdensity field in Eulerian coordinates in terms of the displacement field:

\be
\delta(\mathbf{x})=-\nabla_\v{x}\cdot\v{\Psi}(\v{x}).
\ee 

In Fourier space the displacement field is thus given by:
\be 
\label{eq:psi}
\tilde{\v{\Psi}}(\v{k})=\frac{i\v{k}}{k^2}\tilde{\delta}(\v{k}).
\ee

The idea of BAO reconstruction is to get an estimate of the large scale displacement field from the observed non-linear density field and then use this field to displace the galaxies back to their initial positions. This results in removing the displacements of galaxies on large scales that contribute the most to the smearing of the acoustic peak. When considering also the redshift-space distortions, there are two main ways to do the reconstruction: \textit{anisotropic} and \textit{isotropic} (see \cite{Seo2015a}). In this work we focus on the anisotropic reconstruction in which the redshift-space distortions are kept in the final density field. 
Following the convention of \cite{Seo2015a}, the algorithm proceeds in the following way:
\begin{enumerate}
\item The observed density field is convolved with a smoothing kernel $S(k)$ to reduce the small-scale non-linearities:
$\tilde{\delta}(\v{k})\rightarrow\tilde{\delta}_{\textrm{nl}}^s(\v{k})S(k)$,
where $S(k)$ is usually a Gaussian filter $S(k)=\exp[-0.5k^2R_{\Psi}^2]$ with $R_{\Psi}$ the displacement smoothing scale and $\tilde{\delta}_\textrm{nl}^s$ is the observed redshift-space density field.
\item We estimate the negative real-space displacement field from the smoothed density field:
$$\tilde{\v{s}}^r(\v{k})=-\frac{i\v{k}}{k^2}\frac{\tilde{\delta}_{\textrm{nl}}^s(\v{k})}{b}S(k),$$
where $b$ is the linear galaxy bias.
\item We displace the galaxies by: $$\v{s}^s(\v{x})=\v{s}^r(\v{x})+\frac{f-\beta}{1+\beta}(\v{s}^r(\v{x})\cdot \v{z})\v{z}$$ to obtain the displaced density field $\delta_d(\v{x})$, 
where $f$ is the growth rate and $\beta$ is the redshift-space distortion parameter: $\beta=f/b$ and $\v{s}^{s}(\v{x})$ is the negative redshift-space displacement field.
\item We shift a uniformly distributed grid of particles by the same $\v{s}^s$ to obtain the shifted density field $\delta_s(\v{x}).$
\item The reconstructed density field is then defined as $\delta_r(\v{x})\equiv\delta_d(\v{x})-\delta_s(\v{x}).$
\end{enumerate}

While this method is intended for observations of galaxies in redshift-space, one can also apply it to a particle set such as the matter density field from an N-body simulation by setting $b=1$. If the galaxy/matter catalogue is in real-space, redshift-space distortions can be switched off by setting $\beta=f=0$. When applying this method to halo catalogues, instead of galaxies, we use linear halo bias $b_{\rm{h}}$.

\subsection{Pixelated BAO reconstruction}
\label{subsec:algorithm}

The standard reconstruction improves the significance of the BAO peak position in the power spectrum or the correlation function of the observed galaxy distribution. However, in the description of the algorithm in the previous section, the fact that the density field was estimated from a discrete number of tracers never played any role\footnote{Although note that estimates of the displacement field from very sparse samples affect the performance of BAO reconstruction \cite{White2010}.}. Moreover the ZA, or higher order LPT, are thought to effectively describe the motion of dark matter fluid elements, which could end up containing more than one galaxy. It is therefore worth to see how BAO reconstruction performs on mesh-based fields, and in this section we define the relevant modifications to the original method required when dealing with pixels. A similar method was presented in \cite{Seo2015} to derive the reconstructed density field in the presence of the foreground wedge effect.

The main modification compared to the standard reconstruction technique is that we work at the level of a regular grid and we treat the grid cells in the simulations as galaxies in the standard reconstruction algorithm. The grid cells we use are the same as the ones we used to produce the matter and halo maps, as described in \ref{subsec:maps}. Once we have the matter and halo maps, we proceed to compute the displacement field using the already smoothed density field $\tilde{\delta}_{\rm{sm}}(\v{k})$ We do this by first applying a Gaussian smoothing kernel to $\tilde{\delta}_{\rm{sm}}(\v{k})$ such that it is effectively smoothed isotropically with a displacement smoothing scale $R_{\Psi}$:
\be
\tilde{\delta}_{\v{\Psi}}(\v{k})=\tilde{\delta}_{\rm{sm}}(\v{k})\exp\left[-\frac{k^2}{2}((1-\mu^2)R_\perp^2 + \mu^2 R_{\Psi}^2)\right],
\ee
where $R_\perp=\sqrt{R_{\Psi}^2-\sigma^2}$ is the transverse smoothing scale which is smaller than $R_\Psi$ to take into account the fact that we have already smoothed the field in the transverse direction. The choice of $R_{\Psi}$ will be discussed bellow.

Using this overdensity field we calculate the negative displacement field at the centres of grid cells using:
\be
\v{s}(\v{x})=\textrm{IFFT}\left[-\frac{i \v{k}}{k^2}\tilde{\delta}_{\v{\Psi}}(\v{k})\right].
\ee

We then use this displacement field to move the centres of grid cells according to the derived displacement field. Next, we compute the displaced field $\delta_d(\v{x})$ of displaced grid cells on a regular grid using CIC scheme and weighting each grid cell by $(1+\delta_d)$. When computing the shifted field we only need to modify the weights in the CIC scheme since in both the displaced and the shifted field case the initial grid cells have been displaced by the same displacement field. We thus use the positions of displaced grid cells and apply equal weights using the CIC scheme to compute the shifted field $\delta_s(\v{x})$.

In the last step we subtract shifted field from the displaced field to obtain the reconstructed density field: 
$$\delta_r(\v{x})=\delta_d(\v{x})-\delta_s(\v{x}).$$

In the case of $\sigma=0$, we tested several sizes of grid cells and we find that the reconstruction improves as we increase the resolution, converging when the size of the grid cells approaches the mean particle separation in the simulation. In our case this separation is $1000\,\hMpc/512$ $\approx2\,\hMpc$, and we use this size for the rest of the paper. We also find that the choice of grid cell size that we use for performing reconstruction does not depend on the angular resolution $\sigma$. This is mainly due to the fact that the radial direction is unaffected by the angular resolution and having smaller grid cell sizes along the radial direction improves the reconstructed density field.

\begin{figure}[htbp]
\centering
\subfloat{\includegraphics[scale = 0.39]{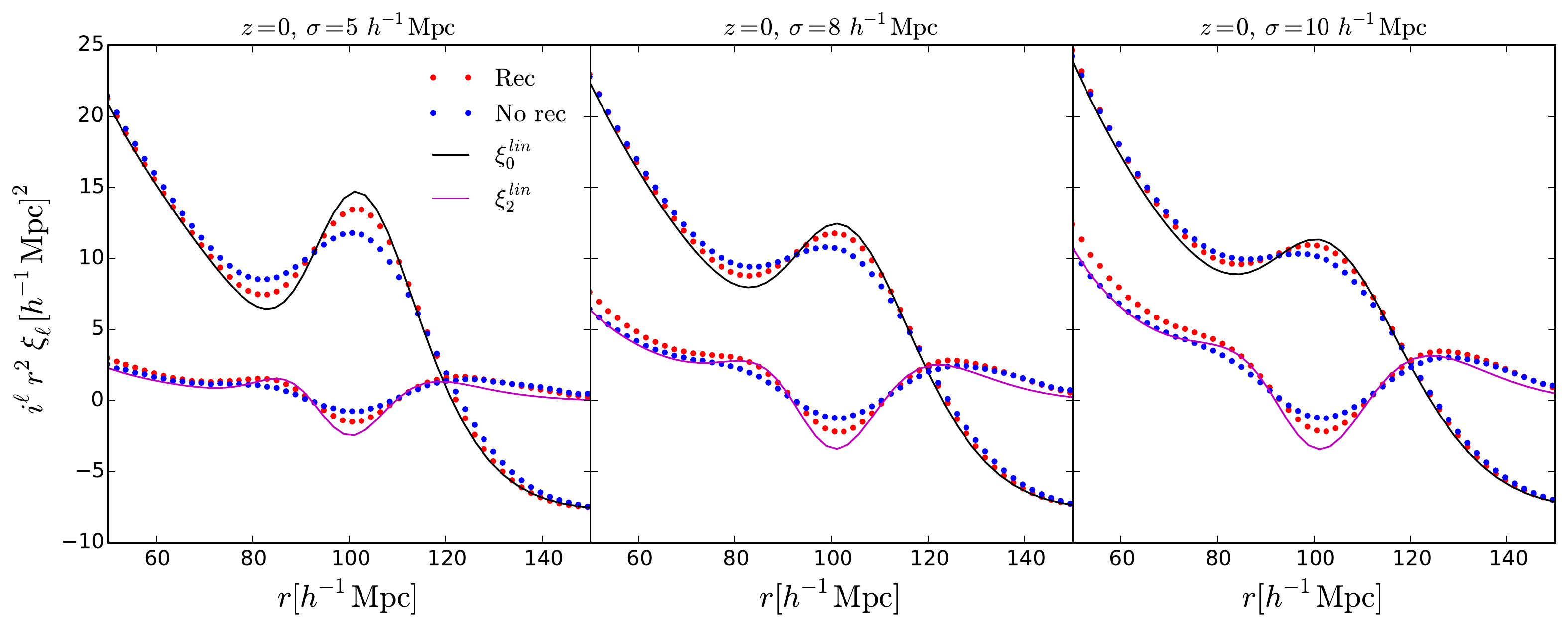}}\\
\subfloat{\includegraphics[scale = 0.39]{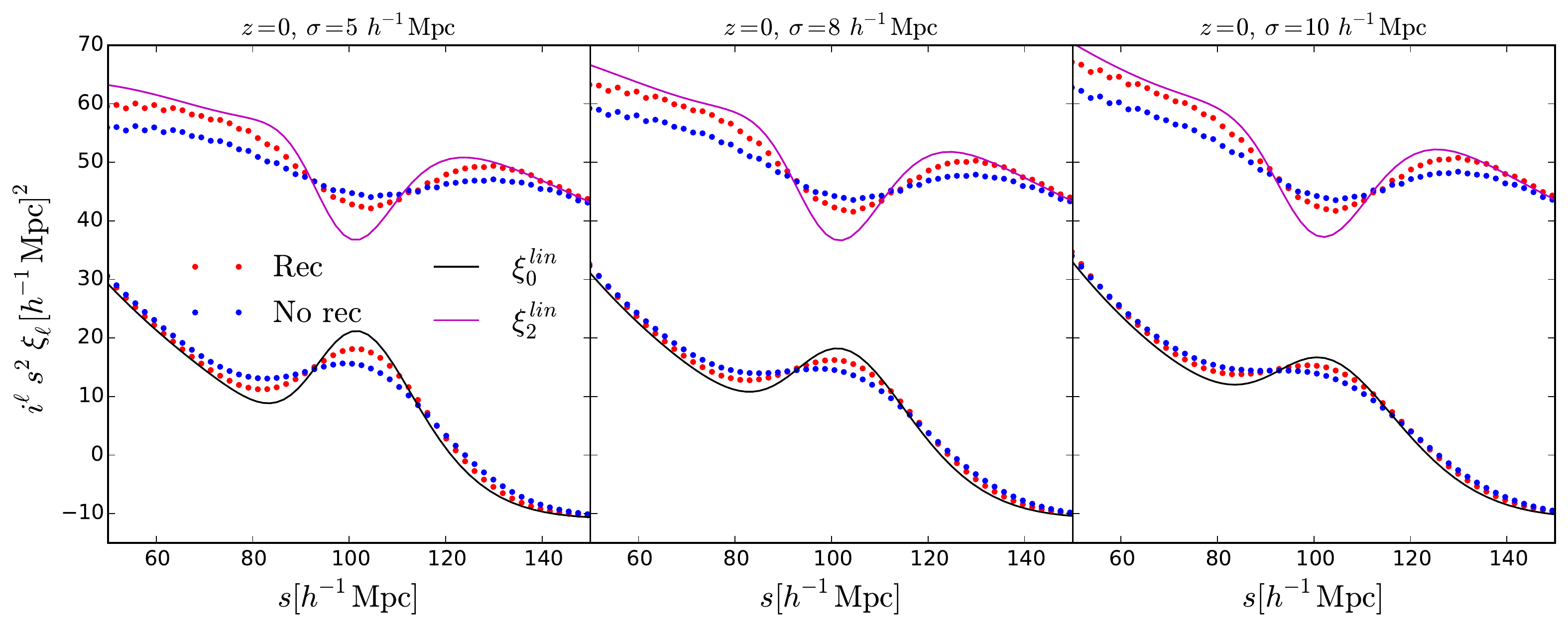}}
\caption{Average monopoles and quadrupoles of the pixelated matter maps in real-space (top) and redshift-space (bottom) at $z=0$ before (blue) and after (red) reconstruction in cases of different angular resolution. Overplotted is the linear theory prediction in solid lines.}
\label{fig:Matter_real_z0_plots}
\end{figure}

\begin{figure}[htbp]
\centering
\subfloat{\includegraphics[scale = 0.4]{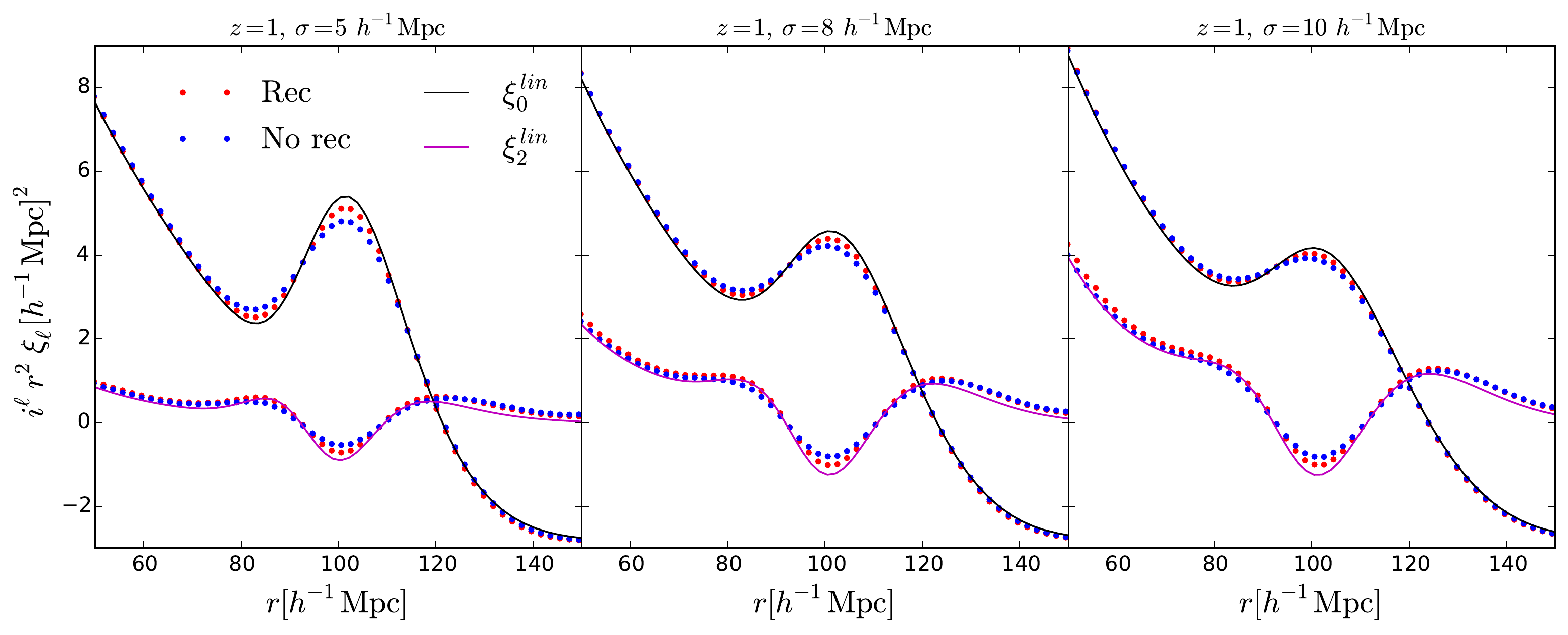}}\\
\subfloat{\includegraphics[scale = 0.4]{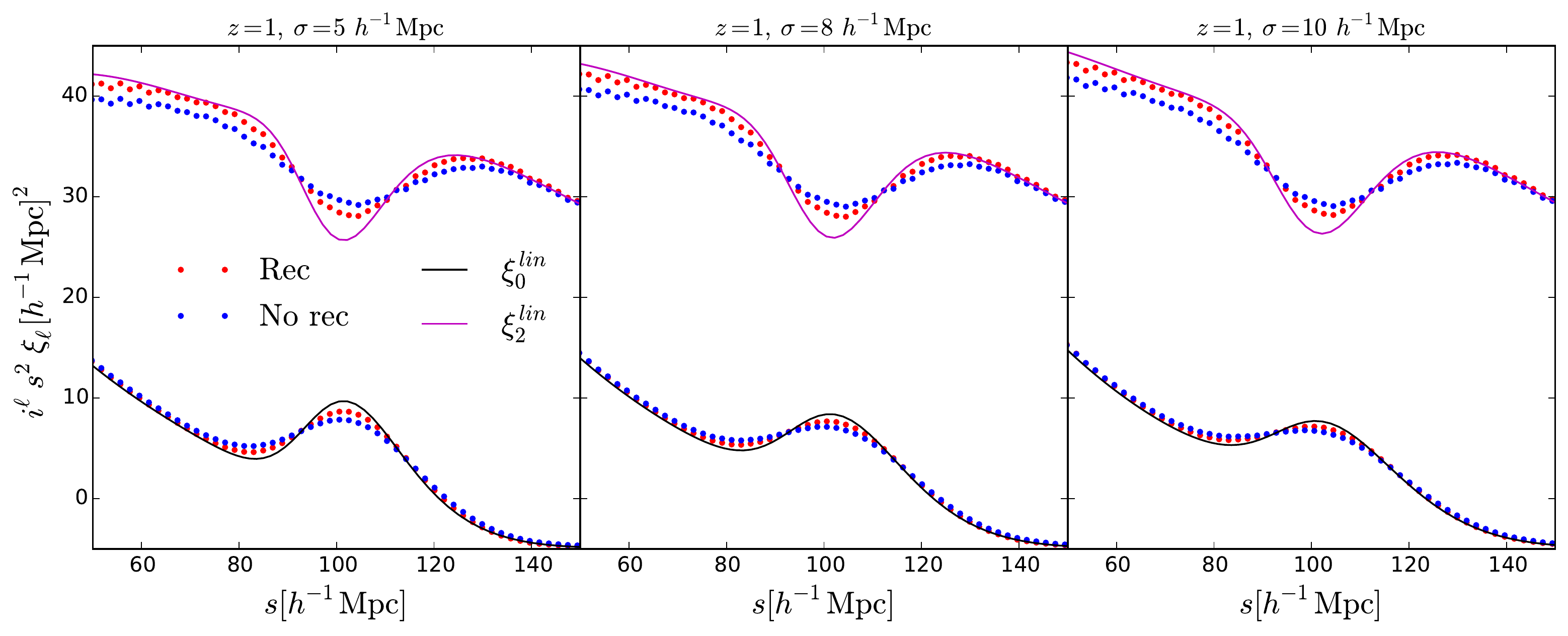}}
\caption{Average monopoles and quadrupoles of the pixelated matter maps in real-space (top) and redshift-space (bottom) at $z=1$ before (blue) and after (red) reconstruction in cases of different angular resolution. Overplotted is the linear theory prediction in solid lines.}
\label{fig:Matter_real_z1_plots}
\end{figure}

\begin{figure}[htbp]
\centering
\subfloat{\includegraphics[scale = 0.4]{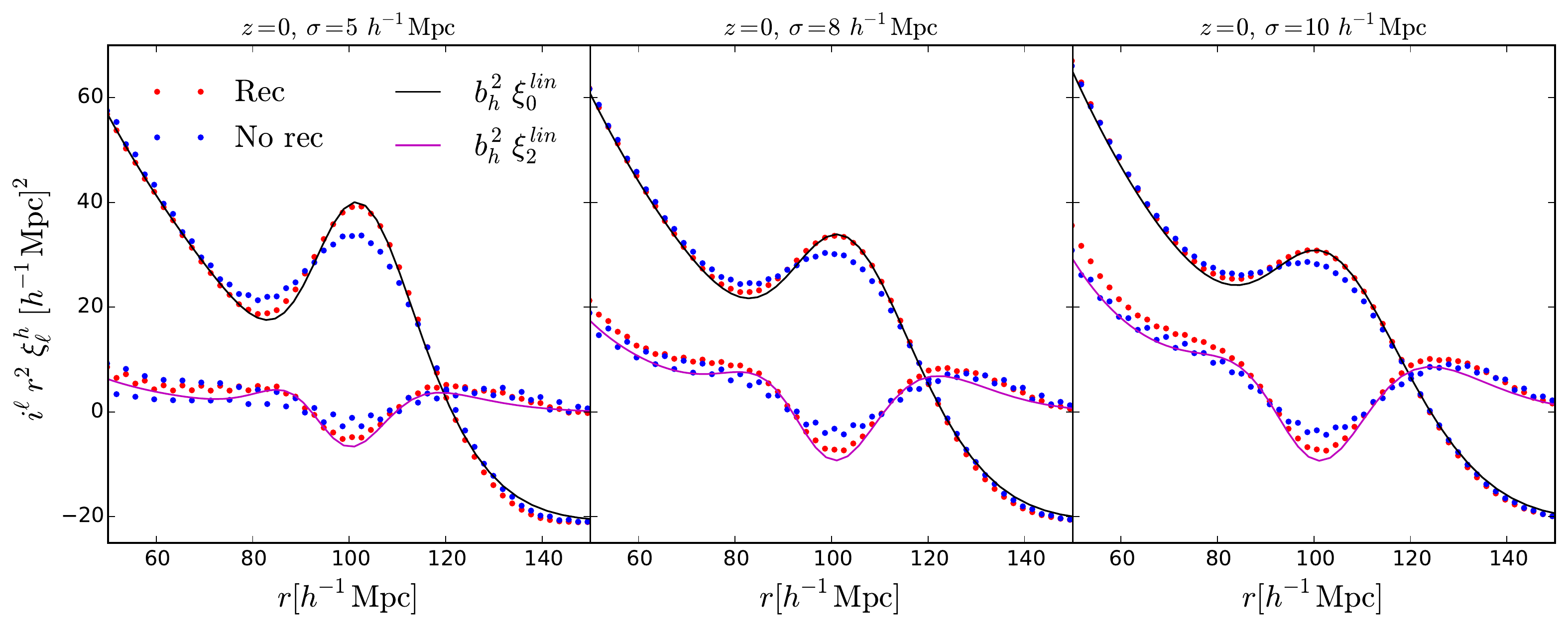}}\\
\subfloat{\includegraphics[scale = 0.4]{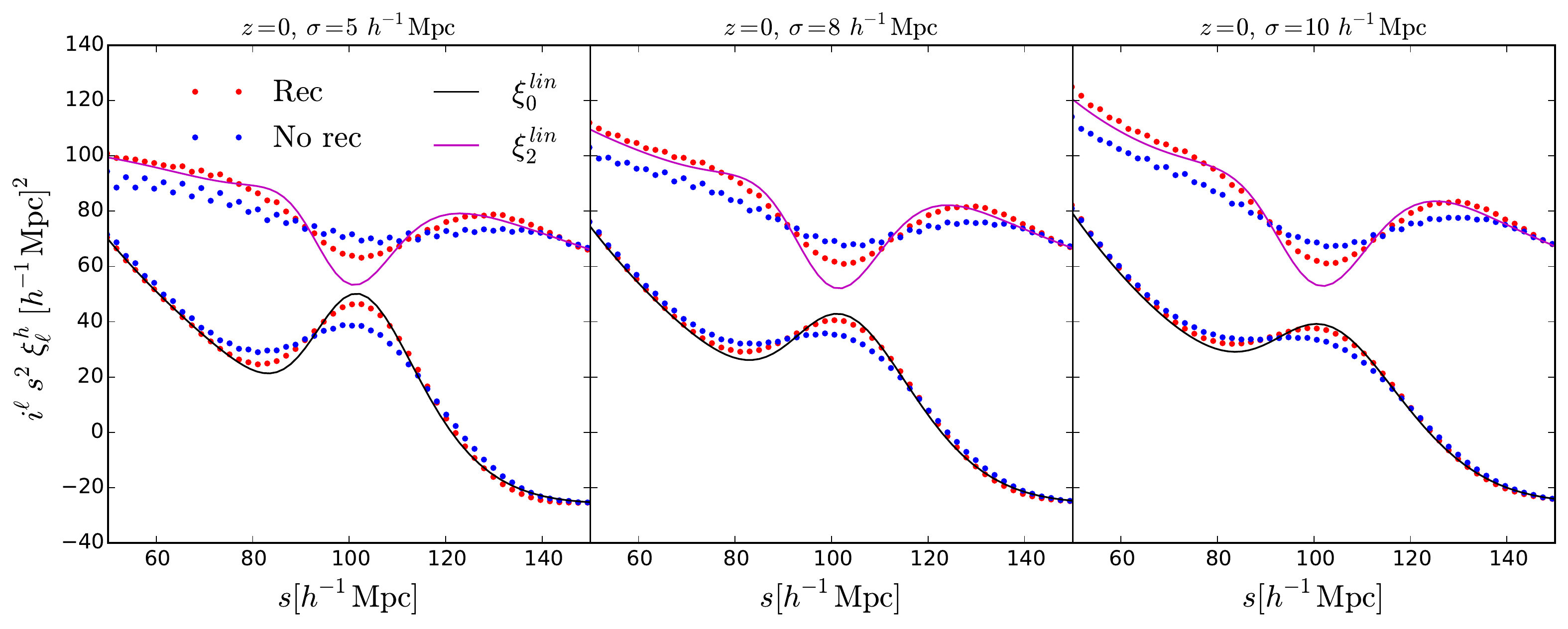}}
\caption{Average monopoles and quadrupoles of the pixelated halos maps in real-space (top) and redshift-space (bottom) at $z=0$ before (blue) and after (red) reconstruction in cases of different angular resolution. Overplotted is the linear theory prediction in solid lines.}
\label{fig:Halos_real_z0_plots}
\end{figure}

We have tested this method using the matter and halo maps in real- and redshift-space at $z=0$ and $z=1$ created from 500 COLA simulations. In Figure \ref{fig:Matter_real_z0_plots} we show the average monopole and quadrupole at $z=0$ in real- and redshift-space before and after reconstruction of the matter maps. We show the results at $z=1$ in Figure \ref{fig:Matter_real_z1_plots}. In Figure \ref{fig:Halos_real_z0_plots} we show the average monopole and quadrupole at $z=0$ in real- and redshift-space before and after reconstruction of the halo maps. We would like to note that in Figures \ref{fig:Matter_real_z0_plots}, \ref{fig:Matter_real_z1_plots} and \ref{fig:Halos_real_z0_plots} the position of the linear point at roughly $90\,\hMpc$, as proposed recently in \cite{Anselmi}, remains unchanged with varying the angular resolution. Perhaps more importantly is that it appears invariant after reconstruction, both in scale and height.

\begin{figure}[htbp]
\centering
\subfloat{\includegraphics[scale = 0.5]{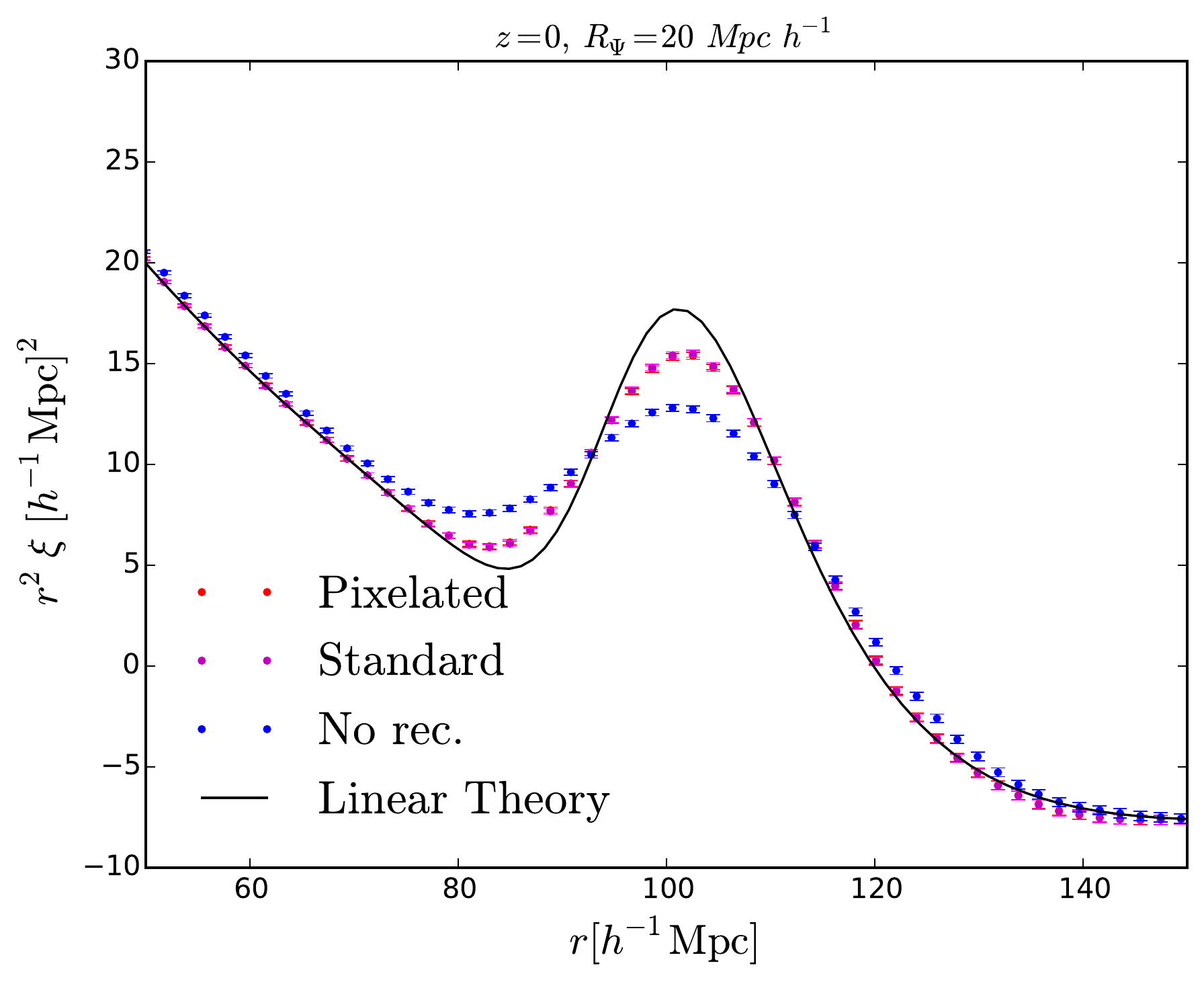}}
\caption{Measured real-space matter density field correlation function at $z=0$ after using the standard reconstruction (magenta) and our method (red) to a set of 500 simulations of matter in real-space. Also shown is the non-reconstructed correlation function (blue) and the linear theory prediction (black).  The data points show the mean values and the error bars show the scatter of the mean values. Standard and our reconstruction method overlap within the error bars of the mean.}
\label{fig:Stand_vs_PM_rec_real_z0}
\end{figure}

We also apply our reconstruction method to matter maps in real-space at $z=0$ that correspond to a galaxy survey ($\sigma=0$). We measure the correlation function in each of the simulations before and after performing both standard and our reconstruction method. In Figure \ref{fig:Stand_vs_PM_rec_real_z0} we show the comparison between the average measured correlation function using the standard and our reconstruction method. In both cases we use the same displacement smoothing scale $R_{\Psi}=20\,\hMpc$. We see that the two methods basically overlap inside the uncertainty on the mean. In section \ref{sec:ST_vs_PM} we show a more quantitative comparison and agreement between the two methods.
In the case of a galaxy survey, we find that this way of performing reconstruction is almost identical to the standard reconstruction as long as the cell sizes are small and the CIC correction is properly accounted for. Furthermore, it is less computationally expensive, since there is no need to 1) interpolate the displacement field for every particle and 2) generate uniform random field of particles, interpolate the displacement field and move the particles.

\subsection{Smoothing scale for the displacement field} \label{subsec:r_psi}
\begin{figure}[htbp]
\centering
\subfloat{\includegraphics[scale = 0.4]{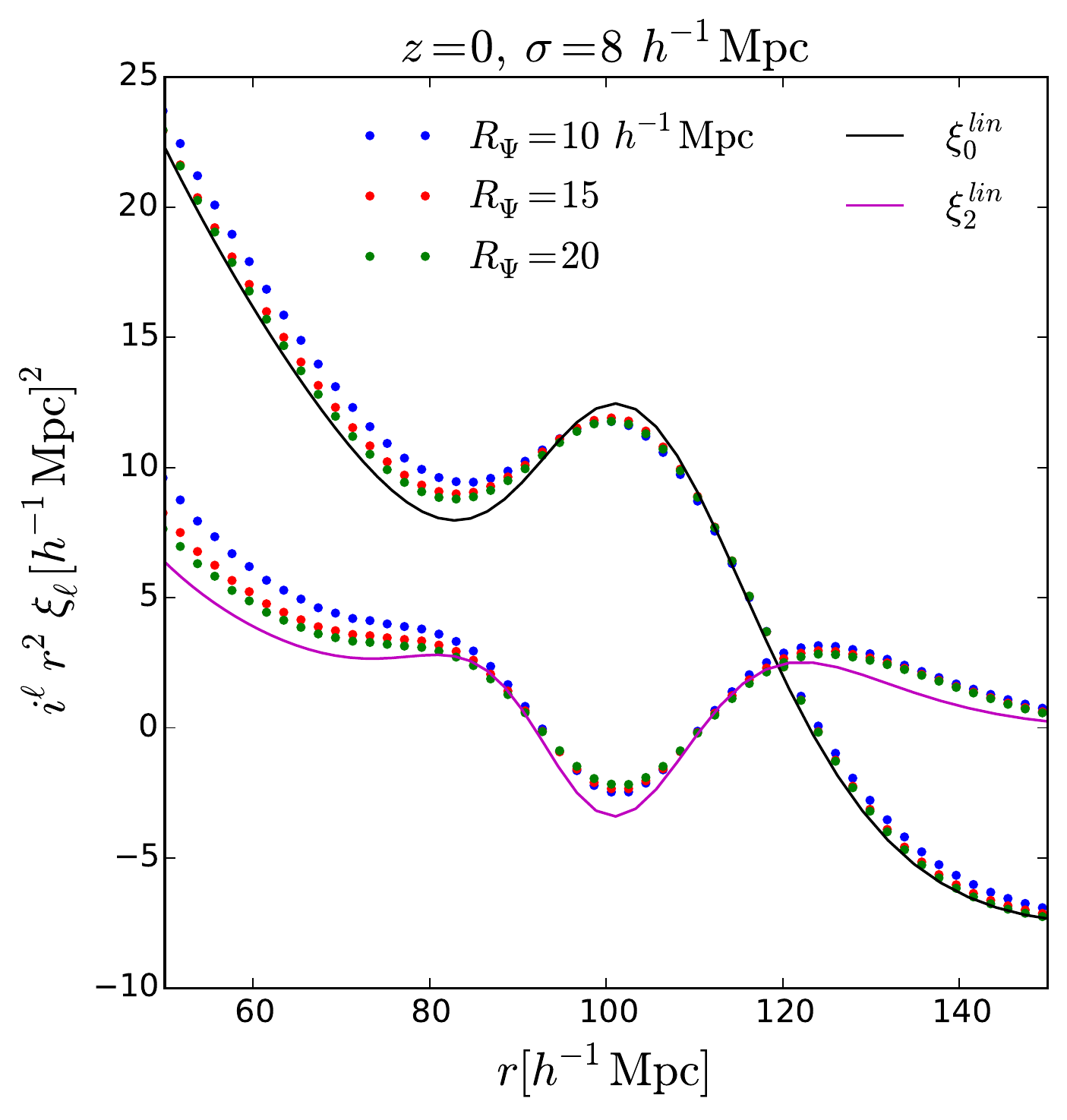}}
\subfloat{\includegraphics[scale = 0.4]{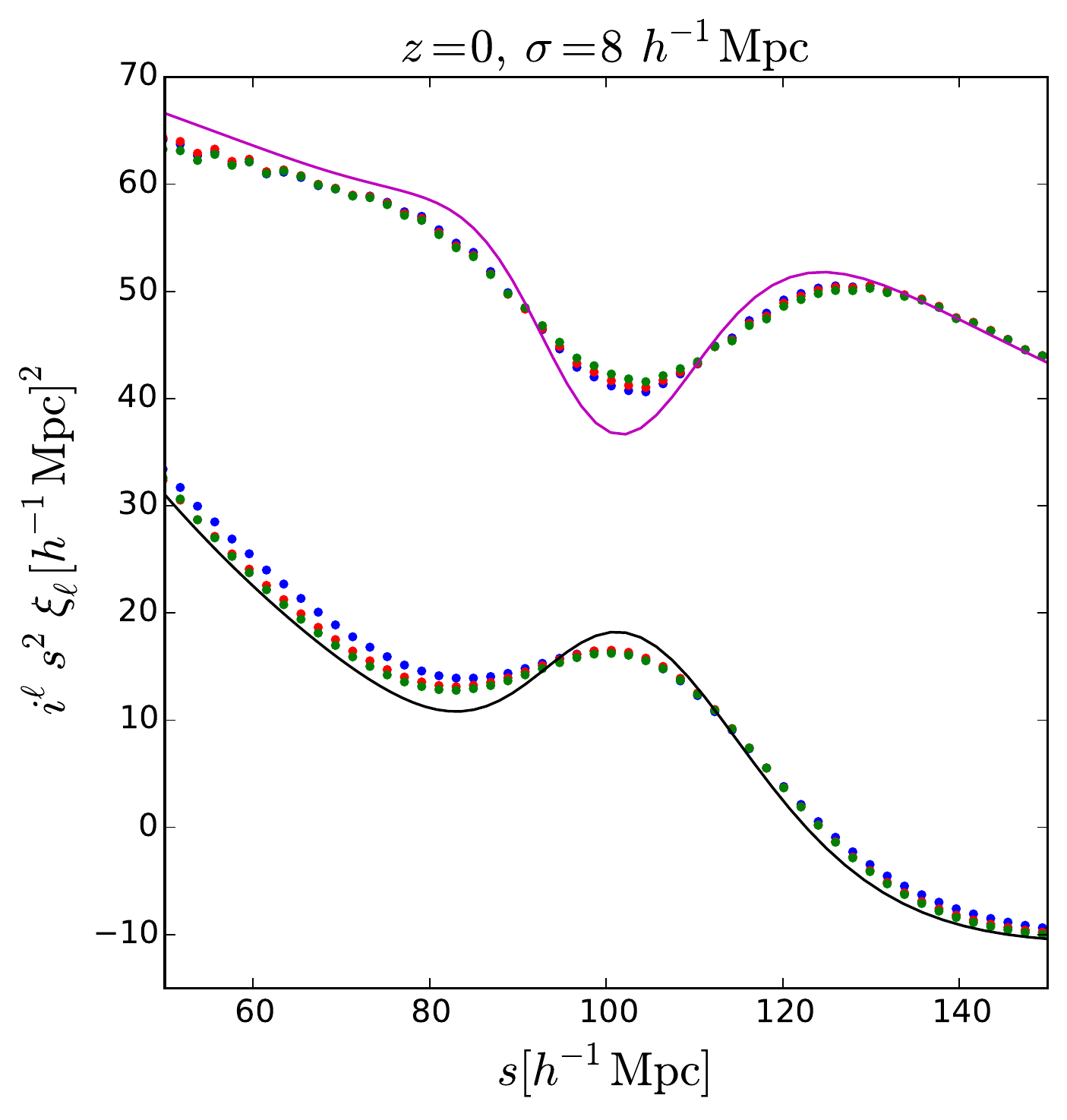}}
\caption{Mean monopoles and quadrupoles of pixelated matter maps in real-space (left) and redshift-space (right) after reconstruction using different smoothing scales for the displacement field: $R_{\Psi}=10,\,15,\, 20\,h^{-1}\textrm{Mpc}$. Here we are only showing results using $\sigma=8\hMpc$ while similar results hold for other values of $\sigma$. }
\label{fig:R_psi_comp}
\end{figure}

The choice of the smoothing scale $R_{\Psi}$ should be made such as to tame the non-linearities at very small scales, while at the same time keep the valuable information of the mildly non-linear regime. The first requirement means making this scale larger, while the second requires it to be not too large. The impact of the smoothing scale $R_{\Psi}$ on the standard reconstruction performance has been previously studied in detail both in mocks and data, e.g.\ \cite{Padmanabhan_2012, Burden2014, Vargas-Magana2015, Seo2015a}. The choice depends on the level of non-linearity in the density field and in the shot-noise contribution \cite{White2010, Seo2015, Cohn2015}. Optimal choice of the scale depends on the case in study and has a broad range of values, ranging from $5-15~\hMpc$. We are facing a somewhat different situation when we study the observables with low angular resolution. Therefore we tested the performance of our reconstruction method using different smoothing scales. In Figure \ref{fig:R_psi_comp} we show mean measured monopole and quadrupole of matter correlation function in real- and redshift-space at $z=0$ after reconstruction for several different smoothing scales $R_{\Psi}$. Using $R_{\Psi}=20\hMpc$ we find better agreement with the linear theory both in real- and redshift-space. We use this value in the rest of the paper and leave the full analysis of the impact of this choice for future work.

\section{Analysis - Fitting the BAO peak}
\label{sec:analysis}

In this section we describe the models we use to obtain the templates for the non-linear correlation functions we are measuring. We then use these templates to build up a model that we use to fit the measured correlation functions in several cases. Our analysis is based on previous BAO analyses, which aim at measuring and put constraints on the position of the BAO peak \cite{Padmanabhan2008, Padmanabhan2012, Xu2013, Anderson2014a}. In the isotropic case, the BAO peak position we measure in the correlation function provides a measure of spherically averaged distance $D_V(z)$. We also need to take into account in our model the fact that, even if our template is a good approximation, assuming a fiducial cosmology can shift the measured BAO peak and therefore affect the distance measurement. This shift can be parametrised by:
\be
\alpha=\frac{D_V(z)/r_d}{D_{V,f}(z)/r_{d,f}}
\ee
where $r_d$ is the sound horizon at the drag epoch. Subscript $f$ corresponds to the assumed fiducial cosmology, while the quantities without subscript refer to the true cosmology. 

Once we have anisotropic clustering, like observations in redshift-space or with angular smoothing, we can measure the BAO position both along the line-of-sight and in the transverse direction, corresponding to separately measuring the Hubble parameter $H(z)$ and the angular diameter distance $D_A(z)$, respectively. In this case, assuming a fiducial cosmology different from the true one will shift the measured BAO position differently along the line-of-sight and transverse directions. To account for this we will follow the method based on~\cite{Padmanabhan2008,Xu2013}. In this formalism the isotropic shift $\alpha$ in the BAO positions is defined as:
\be
\label{eq:alpha}
\alpha=\left[\frac{D_A^2(z)}{D_{A,f}^2(z)}\frac{H_f(z)}{H(z)}\right]^{1/3} \frac{r_{d,f}}{r_d}
\ee
and the anisotropic shift $\epsilon$:
\be
1+\epsilon=\left[\frac{D_{A,f}(z)}{D_A(z)}\frac{H_f(z)}{H(z)}\right]^{1/3}.
\ee

Since we are using numerical simulations with a known cosmology, we expect $\alpha=1$ and $\epsilon=0$. 


\subsection{Isotropic case}
In order to compare the standard (ST) and pixelated (PM) reconstruction methods we need a theoretical model, for the measured matter correlation function in real-space. In real-space the correlation function is isotropic and we use the following template:

\be \label{eq:xi_t_iso}
\xi_t(r)=\int \frac{k^3 d\log k}{2\pi^2}P_{\rm{dw}}(k)j_0(kr),
\ee
where $P_{\rm{dw}}(k)$ is the de-wiggled power spectrum \cite{Eisenstein_2007a}. The de-wiggled power spectrum is designed to account for the damping of the power spectrum due to non-linear effects and is given by:
\be \label{eq:Pdw}
P_{\rm{dw}}(k)=\left[P_{\rm{lin}}(k)-P_{\rm{nw}}(k)\right]\exp\left[-\frac{k^2\Sigma_{\rm{nl}}^2}{2}\right]+P_{\rm nw}(k),
\ee
where $P_{\rm{lin}}(k)$ is the linear theory power spectrum computed using CAMB \cite{CAMB}, $P_{\rm{nw}}(k)$ is the linear power spectrum without the BAO wiggles computed using a fitting formula in \cite{Eisenstein_Hu_1998} and $\Sigma_{\rm{nl}}$ is the Gaussian damping scale.  The final model we use to perform the fit is:
\be
\xi^m(r)=B_0^2\xi_t(\alpha r) + A(r)
\ee
where $A(r)$ is a polynomial function:
\be
A(r)=\frac{a_1}{r^2} + \frac{a_2}{r} + a_3
\ee 
introduced to model effects that modify the broadband shape of the measured correlation function such as redshift-space distortions, halo bias and so on. The term $B_0^2$ controls the overall amplitude of the monopole template and, as in the case of the polynomial coefficients, represents a nuisance parameter.


\subsection{Matter maps} \label{sec:anisotropic}
In case of low angular resolution observables, the correlation function becomes anisotropic even in real-space. We write our template for the 2D smoothed non-linear power spectrum of matter field in real-space as:
\be
P_t(k,\mu)=P_{\rm{dw}}(k,\mu)e^{-k^2(1-\mu^2)\sigma^2}
\ee
where the exponential term represents our 2D smoothing of the density field. At $z=0$ for the matter density field we fix $\Sigma_{\mathrm{nl}}=7.5~h^{-1}\mathrm{Mpc}$ for non-reconstructed model and $\Sigma_{\mathrm{nl}}=4.5~h^{-1}\mathrm{Mpc}$ for the reconstructed model, while at $z=1$ we fix $\Sigma_{\mathrm{nl}}=5~h^{-1}\mathrm{Mpc}$ for non- and $\Sigma_{\mathrm{nl}}=3~h^{-1}\mathrm{Mpc}$ for the reconstructed model. We choose these values based on a best fit to the average measured monopole and quadrupole.

In redshift-space we model the 2D smoothed non-linear power spectrum as:
\be
P_t(k,\mu)=(1+\beta\mu^2)^2F(k,\mu,\Sigma_s)P_{\rm{dw}}(k,\mu)e^{-k^2(1-\mu^2)\sigma^2}.
\ee
The term $(1+\beta\mu^2)^2$ is the Kaiser factor \cite{Kaiser_1987}, that models redshift-space distortions on very large scales. We model the finger-of-God (FoG) effect \cite{Hamilton_1998} using a Gaussian form \cite{Bull2015a}: 

\be
F(k,\mu,\Sigma_s)=e^{-k^2\mu^2\Sigma_s^2},
\ee
where $\Sigma_s$ is the streaming scale describing the dispersion of random peculiar velocities along the line-of-sight direction that washes out the information on small scales. Another form usually used is a Lorentzian \cite{Xu2013} with a streaming scale $\Sigma^\prime_s$ which is different from ours by $\Sigma_s=\Sigma^\prime_s\sqrt{2}$. 

In redshift-space, the non-linear damping is not isotropic anymore. To take the anisotropy into account we use the de-wiggled power spectrum $P_{\rm{dw}}(k,\mu)$ given by \cite{Eisenstein_2007a}:
\be
P_{\rm{dw}}(k,\mu)=\left[P_{\rm{lin}}(k)-P_{\rm{nw}}(k)\right]\exp\left[-\frac{k^2(1-\mu^2)\Sigma_\perp^2+k^2\mu^2\Sigma_\parallel^2}{2}\right]+P_{\rm{nw}}(k).
\ee

Non-linear effects that cause the smearing of the BAO peak are modelled by a Gaussian with damping scale $\Sigma_{\mathrm{nl}}^2=(\Sigma_{\parallel}^2+\Sigma_{\perp}^2)/2$, with components $\Sigma_{\parallel}$ along and $\Sigma_{\perp}$ perpendicular to the line-of-sight. 

We fix the components of the damping scale to the best-fit of the average measured monopole and quadrupole over all simulations. For the non-reconstructed case we set $\Sigma_{\parallel}=(1+f)\Sigma_{\perp}$. At $z=0$ for the matter density field we fix $\Sigma_{\perp}=6.5~\hMpc$, $\Sigma_{\parallel}=9.96~\hMpc$ and $\Sigma_s=4.1\sqrt{2}~\hMpc$. For the reconstructed case we fix $\Sigma_{\perp}=\Sigma_{\parallel}=4~\hMpc$ and $\Sigma_s=3.5\sqrt{2}~\hMpc$.
At $z=1$ for the non-reconstructed case we fix $\Sigma_{\perp}=5~\hMpc$, $\Sigma_{\parallel}=9.39~\hMpc$ and $\Sigma_s=2\sqrt{2}~\hMpc$. For the reconstructed case we fix $\Sigma_{\perp}=\Sigma_{\parallel}=3~\hMpc$ and $\Sigma_s=2\sqrt{2}~\hMpc$.

The power spectrum multipoles of the matter maps templates in real- and redshift-space are given by:
\be
P_{\ell,t}(k)=\frac{2\ell+1}{2}\int_{-1}^1P_t(k,\mu)L_\ell(\mu)d\mu,
\ee
where $L_\ell$ is the Legendre polynomial of order $\ell$. 
The multipoles of the correlation function are then given by:
\be
\xi_{\ell,t}(r)=i^\ell\int \frac{k^3 d\log k}{2\pi^2}P_{\ell,t}(k)j_\ell(kr).
\ee

We use the perturbative expansion in terms of $\alpha$ and $\epsilon$ to construct models for the monopole and quadrupole of the matter correlation function \cite{Xu2013}:
\begin{eqnarray}
\xi_0^m(r) &=& B_0^2\xi_{0,t}(\alpha r) + \frac{2}{5}\epsilon\left[3\xi_{2,t}(\alpha r)+\frac{d\xi_{2,t}(\alpha r)}{d\log(r)}\right] + A_0(r),
\\
\xi_2^m(r) &=& 2B_0^2\epsilon \frac{d\xi_{0,t}(\alpha r)}{d\log(r)}
+\left( 1 + \frac{6}{7}\epsilon \right)\xi_{2,t}(\alpha r)
+\frac{4}{7}\epsilon \frac{d\xi_{2,t}(\alpha r)}{d\log(r)} \nonumber \\
&& + \frac{4}{7}\epsilon \left[ 5\xi_{4,t}(\alpha r) + 
\frac{d\xi_{4,t}(\alpha r)}{d\log(r)} \right] + A_2(r),
\end{eqnarray}
where
\be
A_\ell(r)=\frac{a_{\ell,1}}{r^2} + \frac{a_{\ell,2}}{r} + a_{\ell,3}.
\ee

The polynomials $A_{\ell}(r)$ are standardly added in these analysis to account for systematics and in general to model effects not included in the template as non-linear redshift-space distortions and  scale-dependent bias, that are expected to affect the broadband shape of the correlation function but not the position of the BAO peak. 

When performing the best-fit analysis, we keep the following parameters free: $B_0$, coefficients of the $A_\ell$ polynomials, $\alpha$ and $\epsilon$.


\subsection{Halo maps} \label{sec:halos}
For the analysis of halo maps we restrict ourselves to maps at $z=0$. Due to the low mass resolution in our simulations, halos we identify at $z=1$ are not dense enough tracers of the density field and the shot noise is high enough that we do not see any improvement with reconstruction. At $z=1$ the mean number density of halos is $\bar{n}\approx5\times10^{-5}(h\,\rm{Mpc}^{-1})^3$ which is bellow the limit ($\sim 10^{-4}(h\,\rm{Mpc}^{-1})^3$) at which the standard reconstruction gains saturate \cite{White2010}. The mean number density of halos at $z=0$ is $\bar{n}\approx1.4\times10^{-4}(h\,\rm{Mpc}^{-1})^3$ which is above this limit.

In addition, we find that the measured halo quadrupole in our simulations is dominated by noise both in real- and redshift-space (see Figure \ref{fig:Halos_real_z0_plots}). For these reasons we focus on fitting only the monopole of the correlation function of the halo maps at $z=0$.

We use monopole templates for matter maps in real- and redshift-space. The final model we use to perform the fit to the monopole of the halo maps is:
\be
\xi^m_h(r)=B_0^2\xi_{0,t}(\alpha r) + A(r).
\ee

Before performing a fit to the halo monopole, we normalise our template for the halo monopole to the halo bias $b^2_{\rm{h}}$ measured from the simulations. We measure the halo bias as the ratio of halo and matter power spectrum over 500 simulations and take the average value on large scales.

When fitting the results, we fix the non-linear damping scale $\Sigma_{\rm{nl}}$ to the value we find is the best fit to the average measured monopole from the halo maps. In real-space, for the non-reconstructed case we set $\Sigma_{\rm{nl}}=6.5~\hMpc$. After reconstruction we fix $\Sigma_{\rm{nl}}=2.5~\hMpc$. In redshift-space we set $\Sigma_{\parallel}=\Sigma_{\perp}=\Sigma_{\rm{nl}}$. For the non-reconstructed case we set $\Sigma_{\rm{nl}}=6.5~\hMpc$ and $\Sigma_s=3.5\sqrt{2}~\hMpc$, while after reconstruction we fix $\Sigma_{\rm{nl}}=2.5~\hMpc$ and $\Sigma_s=2.8\sqrt{2}~\hMpc$.

When performing the best-fit analysis, we keep the following parameters free: $B_0$, coefficients of the $A(r)$ polynomial and $\alpha$.

\subsection{Fitting procedure}
We assume that the measured correlation function follows a Gaussian distribution. Thus, finding the best model that describe the data is equivalent to minimizing
\be
\chi^2=(\vec{m}-\vec{d})^TC^{-1}(\vec{m}-\vec{d}),
\ee
where $\vec{m}$ and $\vec{d}$ are vectors containing the values of correlation function of model and data, respectively. In the anisotropic case of matter maps, the vectors of the model and the data contain both the results of the monopole and quadrupole, while in the case of halos we use only the monopole values. $C^{-1}$ is the inverse covariance matrix described below.

We fit the results in the range $50~h^{-1}\mathrm{Mpc}\leqslant r\leqslant150~h^{-1}\mathrm{Mpc}$.
 In the case in which we compare standard and pixelated reconstruction in real-space we use bin sizes of $2~h^{-1}\mathrm{Mpc}$ and we thus employ 51 data points for the correlation function. In the anisotropic case of matter maps we use bin sizes of $4~h^{-1}\mathrm{Mpc}$ and in the fitting range we have 25 data points for monopole and the same for quadrupole, a total of 50 data points. When performing a fit to the monopole of the halo maps, we also use bin sizes  $4~h^{-1}\mathrm{Mpc}$ and we have 25 data points.
 
We minimise the $\chi^2$ and sample the model parameter space using Monte Caro Markov Chain (MCMC) using publicly available code {\tt emcee} \cite{emcee}. We apply a 20\% tophat prior on $\alpha$ and $1+\epsilon$ in order to avoid unphysical values for shift parameters in simulations where the BAO peak is less pronounced. We leave all nuisance parameters free.


\subsection{Covariance matrices} \label{covariance_matrix}
We calculate the covariance matrix directly from the simulations: 
\be
C_{ij}=\frac{1}{N_\textrm{s}-1}\sum_{n=1}^{N_\textrm{s}}[d_n(r_i)-\bar d(r_i)][d_n(r_j)-\bar d(r_j)],
\ee
where $N_\textrm{s}$ is the number of simulations, $d_n(r)$ is a vector containing the values of correlation function calculated from $n$th simulation at radius $r$ and $\bar d(r)$ is the vector containing the mean values of correlation function at radius $r$ over all simulations. In anisotropic case, the vectors $d_n(r)$ and $\bar d(r)$ contain both the monopole and quadrupole of the correlation function, while for halo maps we use only monopole values. 

Since we are using a finite number of simulations, the estimated covariance matrix will be affected by sample noise. This results in a biased estimate for the inverse covariance matrix. This bias can be removed when estimating the inverse covariance matrix by multiplying the inverse estimate by \cite{Hartlap_2007}:
\begin{equation}
C^{-1}=C_{\textrm{original}}^{-1}\frac{N_\textrm{s}-N_\textrm{b}-2}{N_\textrm{s}-1},
\end{equation}
where $N_\textrm{b}$ is the number of bins we are using. In the case of comparing standard and pixelated reconstruction method we have 51 data points. For the halo maps we have 25 data bins and for the matter maps we have 50 data bins: 25 from the monopole and 25 from the quadrupole.

Even with this correction, it has been shown that the noise still affects the constraints of the fitting parameters and this has to be accounted for \cite{Percival2014}. We account for this by multiplying all the measured variances of the fitting parameters by a factor that depends on $N_\textrm{b}$, $N_\textrm{s}$ and the number of fitting parameters $N_\textrm{p}$ (see equation 22 in \cite{Percival2014}).

\section{Results}
\label{sec:results}

In this section we present the constraints we derive, in terms of the position of the BAO peak, before and after reconstruction.
First, we show the results of the comparison between standard reconstruction and our method for a standard galaxy survey, which in our method we simplify to the distribution of pixelated matter in real-space from numerical simulations. Then we show the results of applying our reconstruction method to matter and halos maps in real- and redshift-space.

\subsection{Standard versus pixelated reconstruction method}
\label{sec:ST_vs_PM}

We first present the results of comparing the standard (ST) and pixelated (PM) reconstruction methods when applying them on the spatial distribution of matter in real-space at $z=0$ from our numerical simulations. We have performed the analysis using different smoothing scales for the displacement field $R_{\Psi}$. We choose the non-linear damping parameters $\Sigma_{\rm{nl}}$ for each case by a fit to the average of the measured correlation function. As expected, we find these parameters to be smaller for both reconstruction methods compared to the non-reconstructed case and on average their values decrease by a  $\sim 50\%$ after reconstruction. We also find that the obtained values depend on the smoothing scale for the displacement field used. However, we find no significant difference of the reconstructed $\Sigma_{\rm{nl}}$ between the two reconstruction methods. 

We measure the BAO shift parameter $\alpha$ in each of our 500 simulations before and after reconstruction (using the two different methods). In Table \ref{table:Table_pm_vs_st} we give the summary of best-fit results of the BAO shift parameter $\alpha$ as a function of the smoothing scale for the displacement field $R_{\Psi}$ and used $\Sigma_{\rm{nl}}$ for each case.  

\begin{table}[ht!]
\centering
\begin{tabular}{ccccc}
\hline \hline	
	
$R_{\Psi}$	&	Reconstruction	&	$\Sigma_{\mathrm{nl}}$	&	$\alpha$	&	$\mean{\chi^2}/\mathrm{dof}$ \\ 
 $[\hMpc]$ & method & $[\hMpc]$ & & \\\hline
--	&	No	&	8.0	&	$	1.001	\pm	0.018	$	&	42.1/45	\\ \cline{1-5}
\multirow{2}{*}{10}	&	ST	&	3.4	&	$	1.000	\pm	0.008	$	&	42.5/45	\\ 
	&	PM	&	3.5	&	$	0.999	\pm	0.008	$	&	48.6/45	\\ \cline{1-5}
\multirow{2}{*}{15}	&	ST	&	4.3	&	$	1.000	\pm	0.010	$	&	42.6/45	\\
	&	PM	&	4.4	&	$	0.999	\pm	0.009	$	&	43.4/45	\\ \cline{1-5}
\multirow{2}{*}{20}	&	ST	&	5.0	&	$	0.999	\pm	0.011	$	&	42.4/45	\\
	&	PM	&	5.0	&	$	0.999	\pm	0.011	$	&	42.4/45	\\ \hline \hline

\end{tabular}
\caption{Constraints on the BAO shift parameter (column 4) using the matter density field in real space -- corresponding to a galaxy-survey with $\sigma=0$. Results shown are obtained from simulations without applying reconstruction (first row) and after applying reconstruction (rows 2-7). The non-linear damping parameter $\Sigma_{\mathrm{nl}}$ used in the fit is given in Column 3. Results are shown for different smoothing scales for the displacement field $R_{\Psi}$ (column 1) and the mean $\chi^2/\mathrm{dof}$ is given in column 5. The errors shown for $\alpha$ are the standard deviations over 500 simulations. The actual errors on the means of the parameter $\alpha$ are the listed values divided by $\sqrt{500}$.}
\label{table:Table_pm_vs_st}
\end{table}

\begin{figure}[htbp]
\centering
\subfloat{\includegraphics[scale = 0.3]{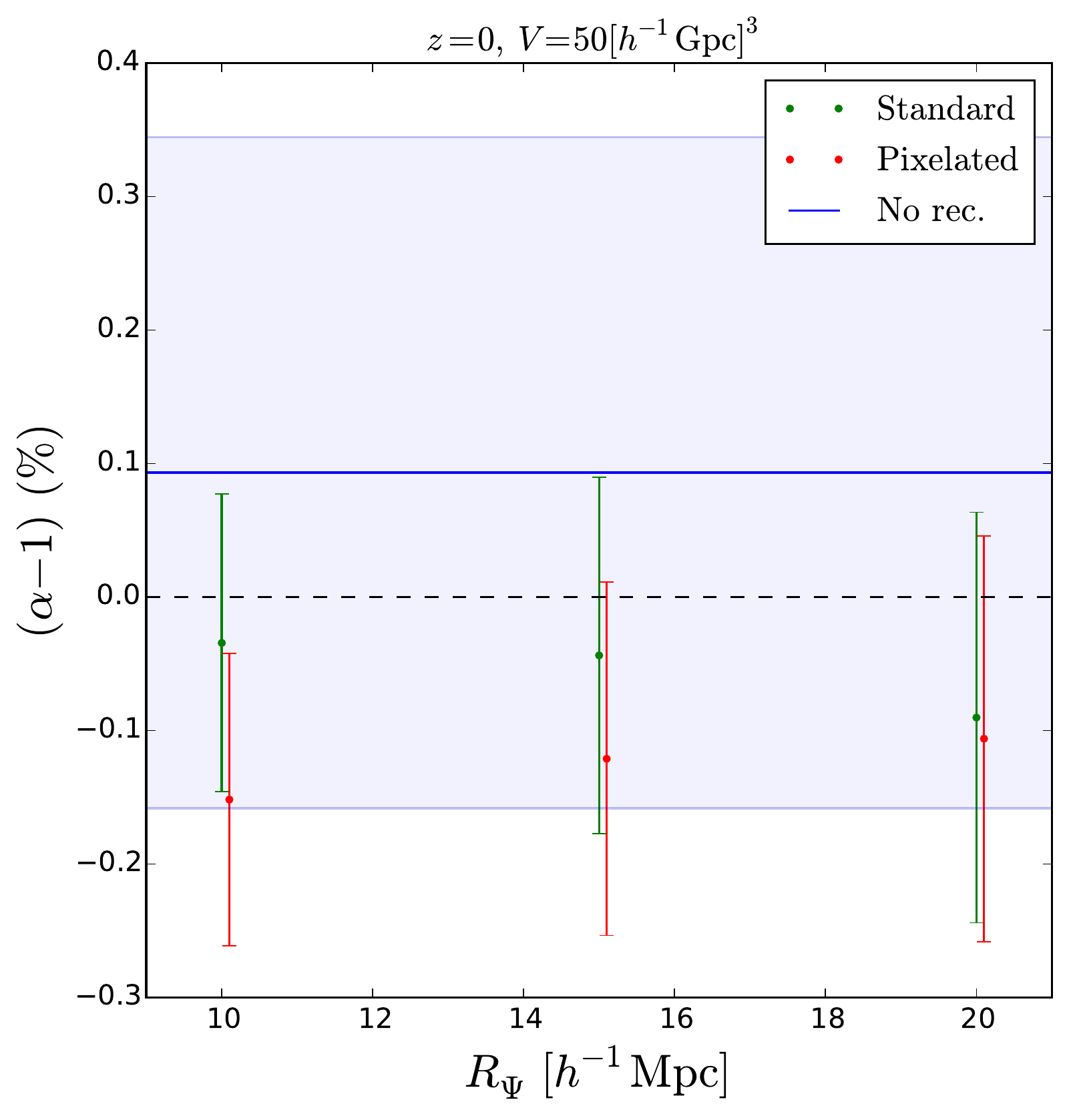}}
\subfloat{\includegraphics[scale = 0.3]{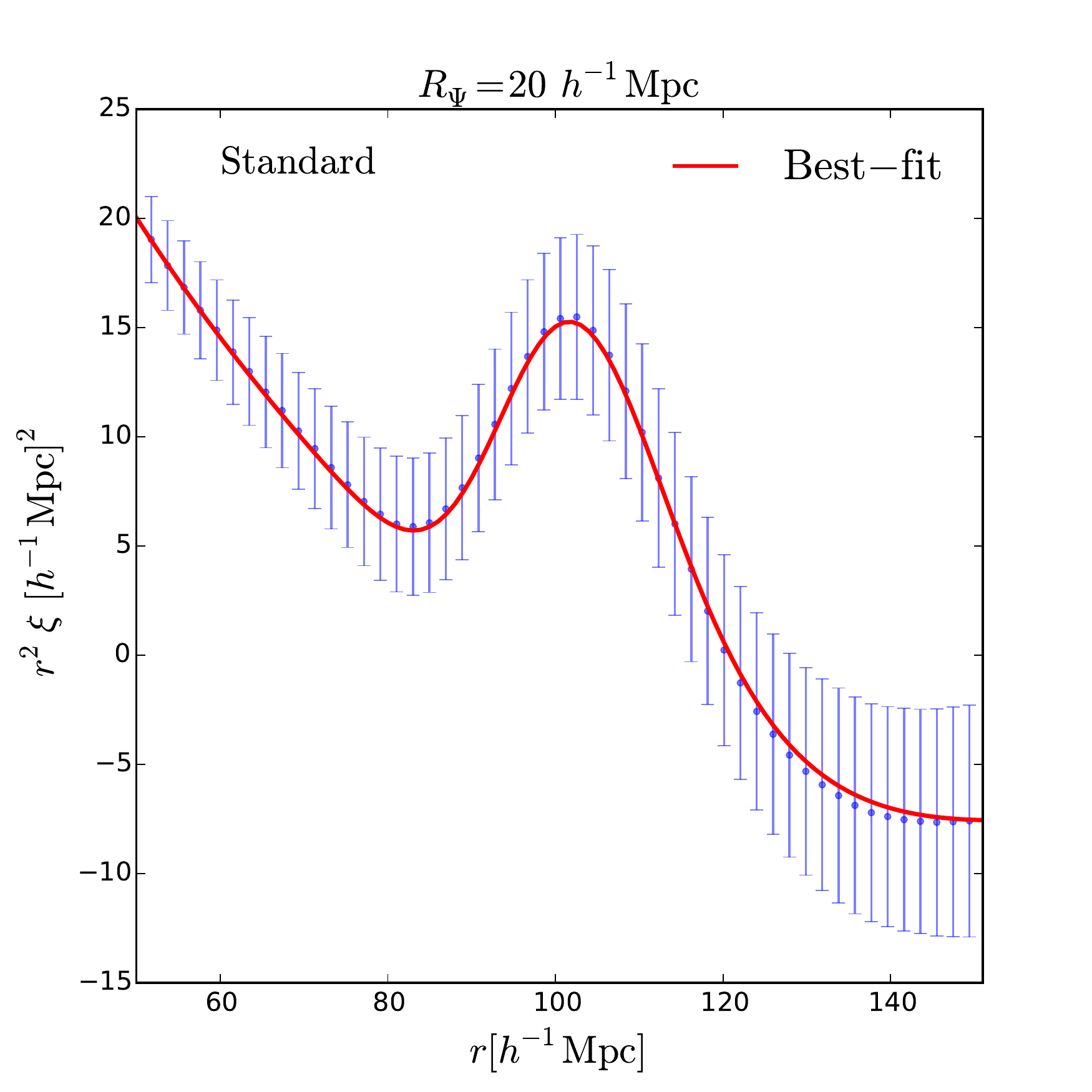}}
\subfloat{\includegraphics[scale = 0.3]{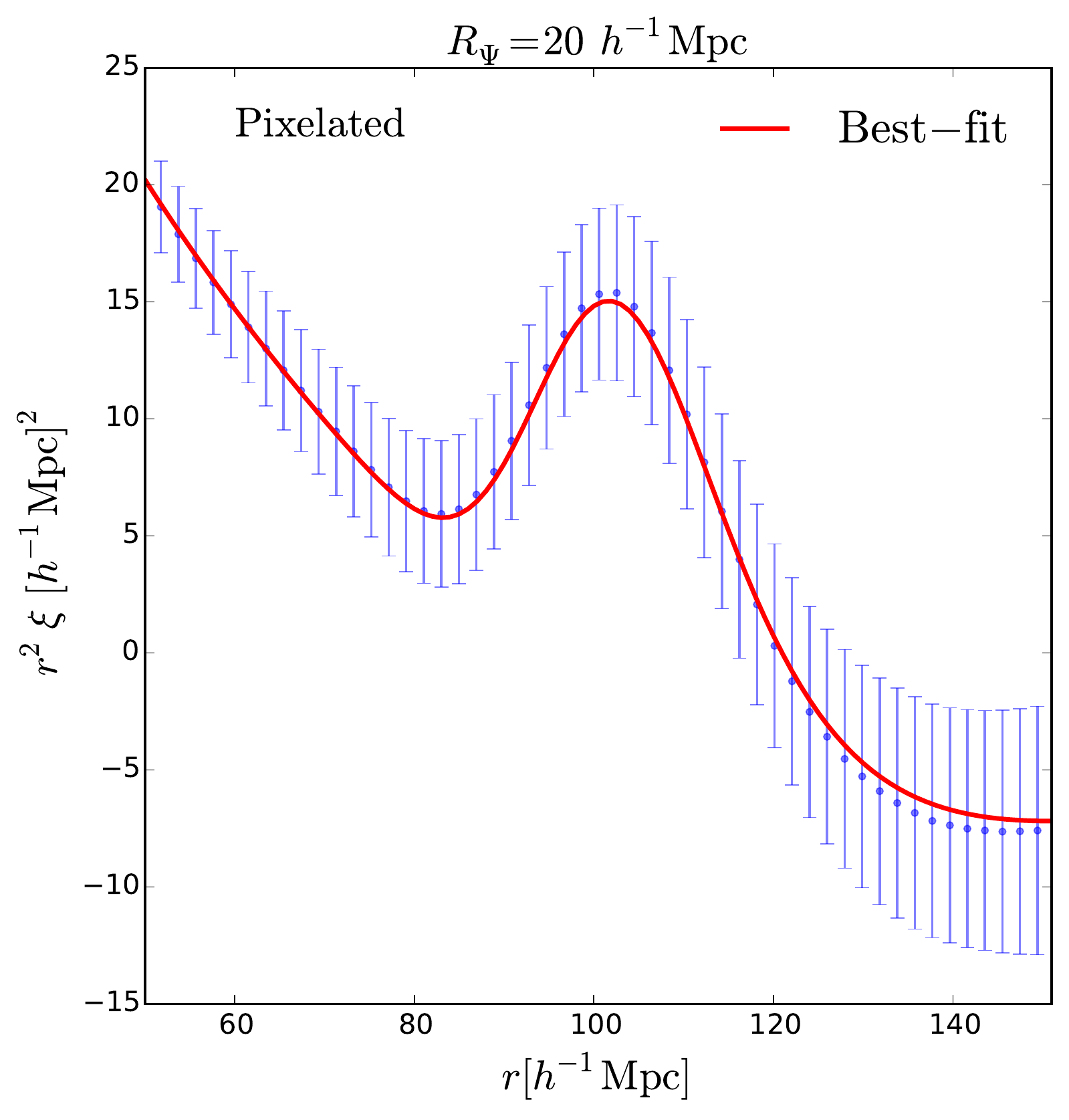}}

\caption{Reconstruction of the matter density field in real space -- corresponding to a galaxy-survey with $\sigma=0$, at $z=0$. \textit{Left panel}: Mean best-fit values of BAO shift parameters $\alpha$ as a function of smoothing scale of the displacement field $R_{\Psi}$ showing the comparison between standard and pixelated reconstruction algorithm. The error bars shown are the standard deviation from 500 simulations divided by $\sqrt{50}$ (i.e. the expected error for a survey covering a volume of $\sim 50~(h^{-1}{\rm Gpc})^3$). \textit{Middle and right panel}: The best-fits to the reconstructed matter correlation function in case of standard and pixelated reconstruction method, using $R_{\Psi}=20\hMpc$.}
\label{fig:Alpha_pm_vs_st}
\end{figure}

In Figure \ref{fig:Alpha_pm_vs_st} (left panel) we plot these results for various smoothing scales for the displacement field $R_{\Psi}$. The points shown are the best-fit values and the error bars shown are standard deviations divided by $\sqrt{50}$ to show the expected uncertainty in a survey of $50\,[h^{-1}\rm{Gpc}]^3$ volume size. Best fits to the reconstructed correlation function in ST and PM case are shown in middle and right panel of Figure \ref{fig:Alpha_pm_vs_st}, respectively, in the case of $R_{\Psi}=20\,\hMpc$. 

We find the obtained values of $\alpha$ to be consistent with the expected value $\alpha=1$ at the level of uncertainties in both ST and PM methods. The uncertainty on $\alpha$ after reconstruction decreases by $40-60\%$, depending on which $R_{\Psi}$ is used. We find no significant difference in ST and PM methods. 


\subsection{Matter maps}

We now focus our attention on the case of matter maps in both real- and redshift-space, considering maps with different resolutions, $\sigma$.
We measure the BAO shift parameters $\alpha$ and $\epsilon$ in each of our 500 simulations before and after reconstruction using the template models outlined in Section \ref{sec:anisotropic}. A summary of the best-fit results both in real- and redshift-space at $z=0$ and $z=1$ is presented in Table \ref{table:Table_matter}. 

\begin{table}[ht!]
\centering
\begin{tabular}{cccccc}

\hline \hline
\multicolumn{6}{c}{Matter maps -- real-space}\\
\hline	
\multirow{8}{*}{$z=0$} & $\sigma[\hMpc]$	&	Reconstruction	&	$\alpha$	&	$\epsilon$	&	$\mean{\chi^2}/\mathrm{dof}$\\ \hline
& \multirow{2}{*}{5}	&	no	&	$1.004 \pm  0.018$	&	$0.001 \pm 0.022$	&	45.3/41\\
&	&	yes	&	$1.0005 \pm  0.0096$	&	$0.002 \pm 0.010$	&	44.1/41\\ \cline{2-6}
& \multirow{2}{*}{8}	&	no	&	$ 1.002\pm  0.021$	&	$0.003 \pm 0.023$	&	45.5/41\\
&	&	yes	&	$ 1.0008\pm  0.0096$	&	$0.003 \pm 0.010$	&	46.2/41\\ \cline{2-6}
& \multirow{2}{*}{10}	&	no	&	$ 1.004\pm 0.019$	&	$0.004 \pm 0.021$	&	45.4/41\\
&	&	yes	&	$ 1.002\pm 0.010$	&	$0.004 \pm 0.011$	&	47.8/41\\ \hline
\multirow{6}{*}{$z=1$} & \multirow{2}{*}{5}	&	no	&	$1.0000 \pm  0.0098$	&	$-0.001 \pm 0.012$	&	43.8/41\\
&	&	yes	&	$0.9983 \pm  0.0070$	&	$0.0008 \pm 0.0080$	&	41.8/41\\ \cline{2-6}
& \multirow{2}{*}{8}	&	no	&	$ 1.003\pm  0.011$	&	$-0.001 \pm 0.013$	&	45.1/41\\
&	&	yes	&	$ 0.9999\pm  0.0071$	&	$0.0006  \pm 0.0088$	&	44.7/41\\ \cline{2-6}
& \multirow{2}{*}{10}	&	no	&	$ 1.005\pm 0.011$	&	$0.001 \pm 0.013$	&	46.3/41\\
&	&	yes	&	$ 1.0021\pm 0.0095$	&	$0.0003 \pm 0.0095$	&	50.4/41\\ \hline

\multicolumn{6}{c}{Matter maps -- redshift-space}\\ \hline

\multirow{8}{*}{$z=0$} & $\sigma[\hMpc]$ &	Reconstruction	&	$\alpha$	&	$\epsilon$	&	$\mean{\chi^2}/\mathrm{dof}$\\ \hline
&	\multirow{2}{*}{5}	&	no	&	$	1.000	\pm	0.023	$	&	$	0.0021	\pm	0.0075	$	&	39.2/41	\\
&		&	yes	&	$	0.998	\pm	0.011	$	&	$	0.0002	\pm	0.0042	$	&	40.0/41	\\ \cline{2-6}
&	\multirow{2}{*}{8}	&	no	&	$	1.001	\pm	0.025	$	&	$	0.0045	\pm	0.0097	$	&	38.6/41	\\
&		&	yes	&	$	0.996	\pm	0.013	$	&	$	0.0011	\pm	0.0060	$	&	40.4/41	\\ \cline{2-6}
&	\multirow{2}{*}{10}	&	no	&	$	1.001	\pm	0.026	$	&	$	0.005	\pm	0.012	$	&	38.9/41	\\
&		&	yes	&	$	0.995	\pm	0.015	$	&	$	0.0016	\pm	0.0076	$	&	41.8/41	\\ \cline{2-6} \hline

\multirow{6}{*}{$z=1$} &	\multirow{2}{*}{5}	&	no	&	$	0.997	\pm	0.014	$	&	$ 0.0012	\pm	0.0077	$	& 30.4/41\\
&		&	yes	&	$	1.000	\pm	0.011	$	&	$	0.0000	\pm	0.0040	$	&	31.9/41	\\ \cline{2-6}
&	\multirow{2}{*}{8}	&	no	&	$	0.998	\pm	0.018	$	&	$	0.001	\pm	0.010	$	&	30.2/41	\\
&		&	yes	&	$	1.000	\pm	0.013	$	&	$	0.0000	\pm	0.0055	$	&	31.6/41	\\ \cline{2-6}
&	\multirow{2}{*}{10}	&	no	&	$	1.000	\pm	0.022	$	&	$	0.000	\pm	0.013	$	&	31.2/41	\\
&		&	yes	&	$	1.001	\pm	0.015	$	&	$	0.0000	\pm	0.0067	$	&	32.4/41	\\ \cline{2-6}
\hline \hline

\end{tabular}
\caption{Constraints on BAO shift parameters $\alpha$ and $\epsilon$ for matter maps with different angular resolutions (column 2) before and after reconstruction. Columns 3 and 4 show the mean and the standard deviation of BAO shift parameters $\alpha$ and $\epsilon$, respectively. The mean $\chi^2/\mathrm{dof}$ is given in column 5. The errors shown for $\alpha$ and $\epsilon$ are the standard deviations over 500 simulations. The actual errors on the means of the parameters $\alpha$ and $\epsilon$ are the listed values divided by $\sqrt{500}$.}
\label{table:Table_matter}
\end{table}

\begin{figure}[htbp]
\centering
\includegraphics[scale = 0.5]{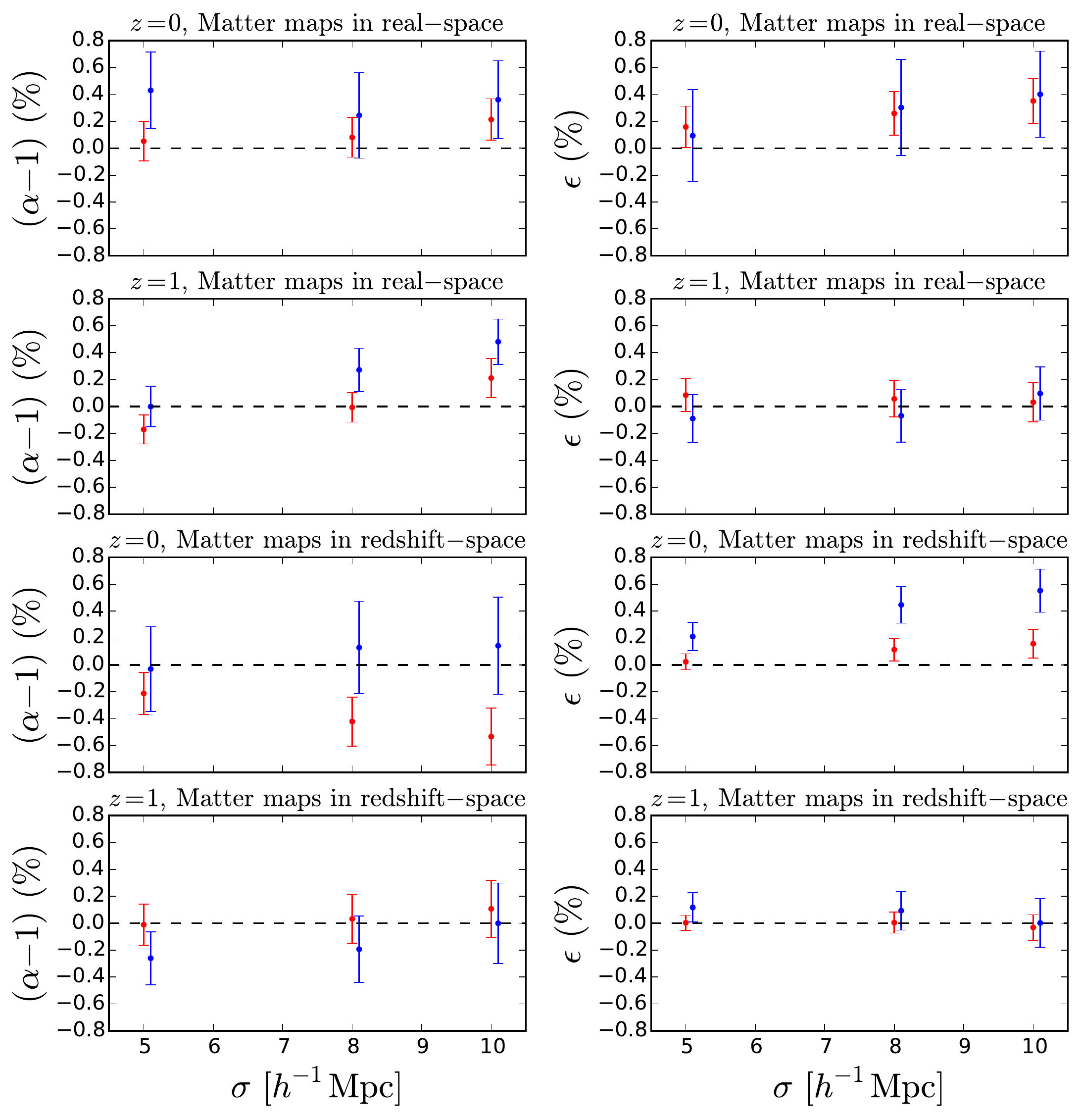}
\caption{Mean best-fit values of BAO shift parameters $\alpha$ (left) and $\epsilon$ (right) as a function of map spatial resolution, $\sigma$, at $z=0$ and $z=1$ for matter maps. Blue points represent the non-reconstructed, while red points represent the reconstructed density field results. The error bars shown are the standard deviation from 500 simulations divided by $\sqrt{50}$ to show the uncertainty we would expect in a survey of $\sim50\,[h^{-1}\rm{Gpc}]^3$ volume size}
\label{fig:Alpha_epsilon_master}
\end{figure}

In Figure \ref{fig:Alpha_epsilon_master} we show the best-fit BAO shift parameters $\alpha$ (left) and $\epsilon$ (right) in real- and redshift-space as a function of the smoothing scale $\sigma$ at $z=0$ and $z=1$. Blue points correspond to non- and red to reconstructed maps fits. The error bars correspond to standard deviations of BAO shift parameters divided by $\sqrt{50}$ to show the expected uncertainty we would expect in a survey of $\sim50\,[h^{-1}\rm{Gpc}]^3$ volume size.

\subsubsection{Real-space}

We find that the uncertainties in both $\alpha$ and $\epsilon$ decrease by $\sim 50\,\%$ after reconstruction across the considered range of map angular resolutions $\sigma$. The 2D smoothing is smearing the BAO peak in the monopole with increasing $\sigma$ so one would expect the uncertainty in $\alpha$ to increase as well. On the other hand, larger smoothing scales make the quadrupole more pronounced and therefore more constraining for the shift parameters. 
After reconstruction, the recovered values of $\alpha$ are closer and consistent with the expected $\alpha=1$ at the single simulation level. Even by considering the error on the mean, i.e. combining the results of all the simulations to probe a volume equal to $500~(h^{-1}{\rm Gpc})^3$, we find that $\alpha$ is consistent with 1 at $5\sigma$ level. We however find a $\sim0.2\%$ shift in the recovered values of $\epsilon$. 

Similar to $z=0$, at $z=1$ we again find that the uncertainties in both $\alpha$ and $\epsilon$ decrease after reconstruction. These decreases are however smaller compared to $z=0$, and reach $\sim 35\%$ in $\alpha$ and $\epsilon$. This is expected since the non-linear effects are smaller at higher redshifts. Recovered means of reconstructed $\alpha$ are within 0.2\% of the expected values, while the reconstructed $\epsilon$ stays within 0.1\% of the expected value and within the uncertainties of the full simulated volume.

\begin{figure}[htbp]
\centering
\includegraphics[scale = 0.3]{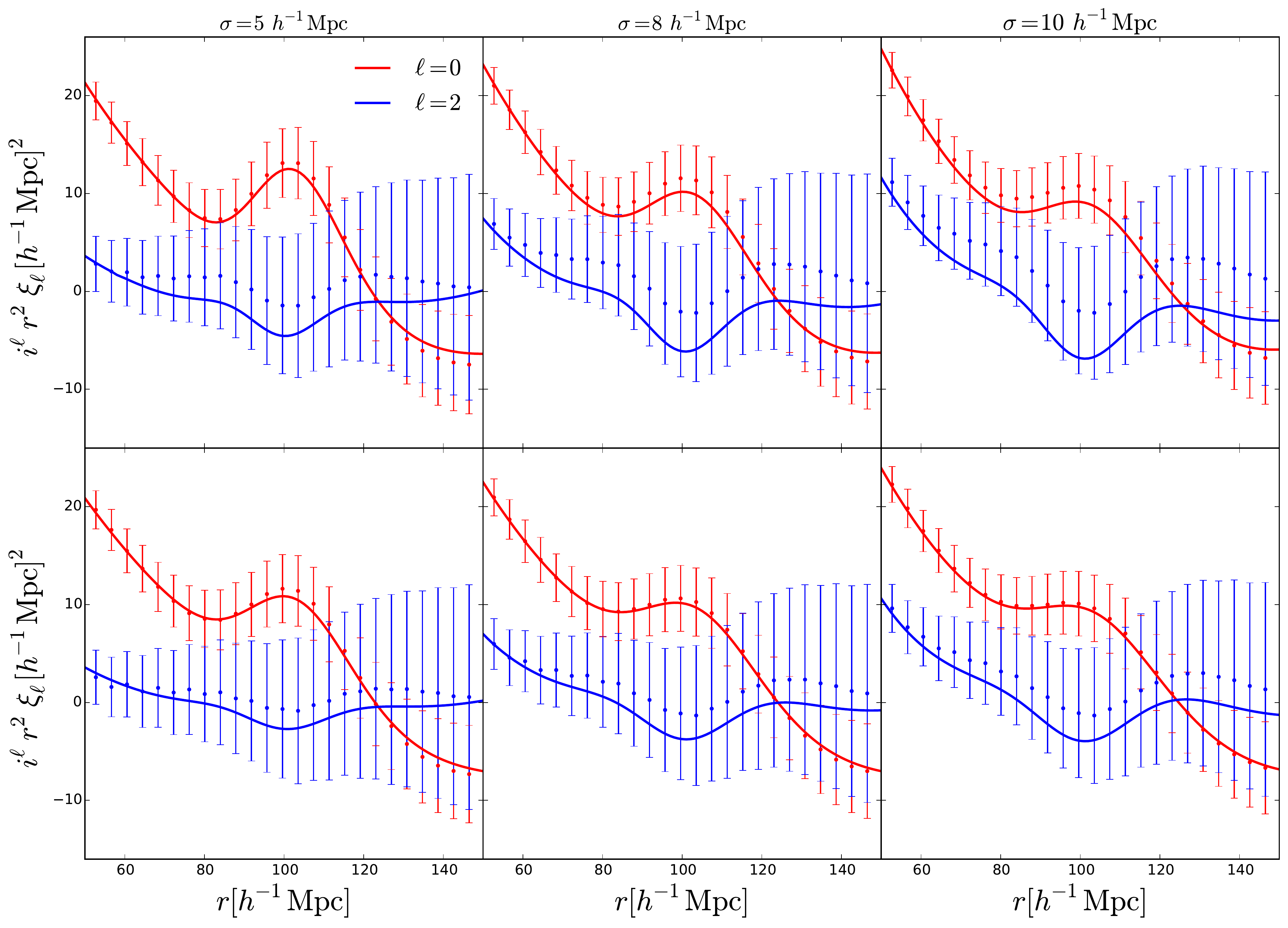}
\caption{The best-fit to monopole and quadrupole correlation function for matter maps in real-space at $z=0$ --  reconstructed (top) and unreconstructed (bottom).}
\label{fig:Best_fit_matter_real_z0}
\end{figure}

\begin{figure}[htbp]
\centering
\includegraphics[scale = 0.3]{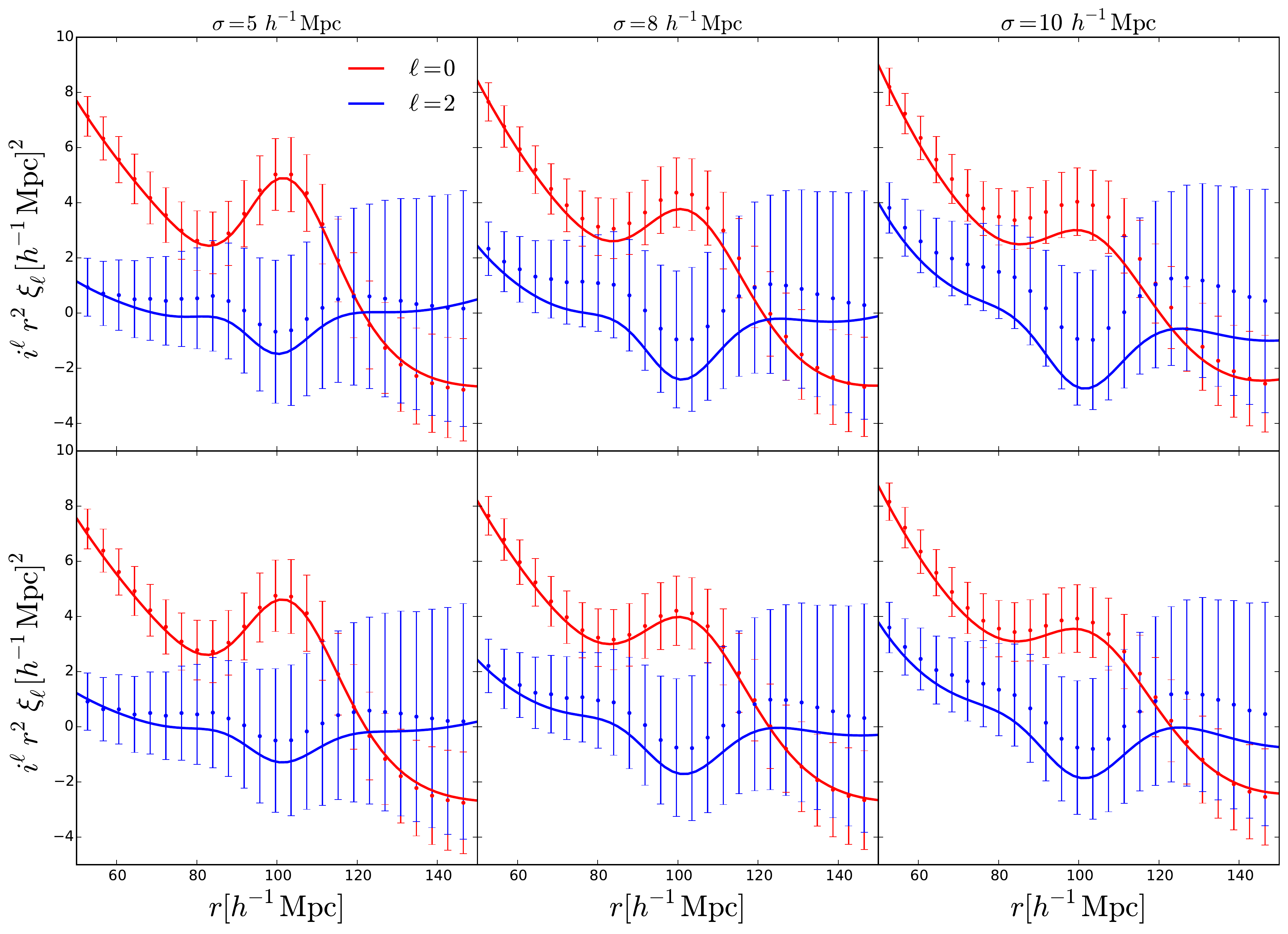}
\caption{The best-fit to monopole and quadrupole correlation function for matter maps in real-space at $z=1$ --  reconstructed (top) and unreconstructed (bottom).}
\label{fig:Best_fit_matter_real_z1}
\end{figure}

In Figures \ref{fig:Best_fit_matter_real_z0} and \ref{fig:Best_fit_matter_real_z1} we show the best-fit to the monopole and quadrupole of the matter maps in real-space after (\textit{top}) and before (\textit{bottom}) reconstruction at $z=0$ and $z=1$, respectively. Comparing the monopole and quadrupole before and after reconstruction, we find the acoustic peak gets more pronounced after reconstruction. This is more evident at $z=0$ than at $z=1$ since the non-linear effects are smaller at higher redshift. We also find that for smaller smoothing scales reconstruction makes the BAO peak more pronounced both in monopole and quadrupole. This improvement of reconstruction decreases as we move to lower resolution maps. Still, one should compare the reconstructed results with the linear theory prediction (solid lines in Figure \ref{fig:Matter_real_z0_plots}, upper panel) and notice that even the linear theory prediction monopole is getting less pronounced as we move to larger values of smoothing scales.

\begin{figure}[htbp]
\centering
\includegraphics[scale = 0.3]{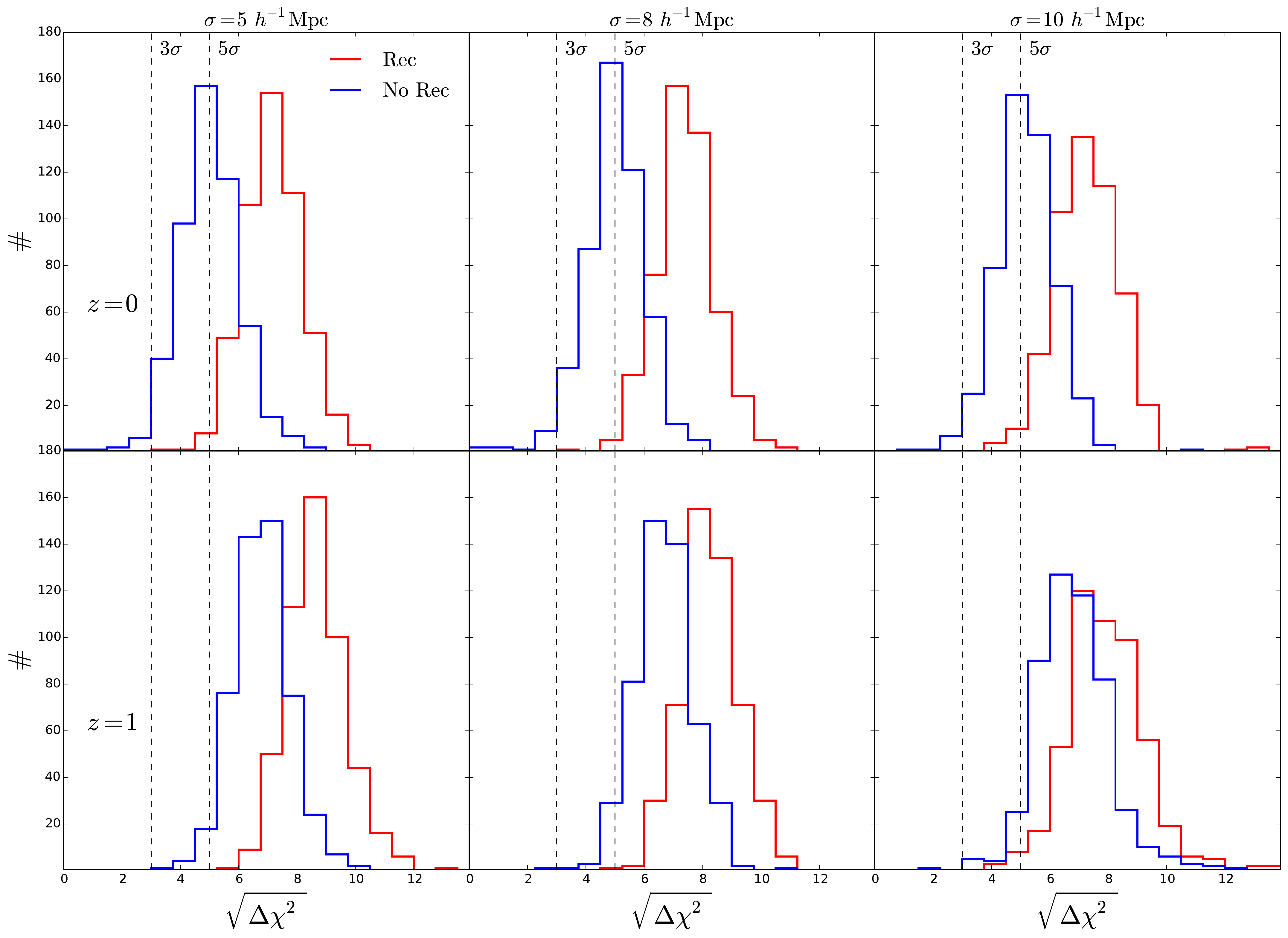}
\caption{Significance of detecting a BAO peak with (red) and without (blue) reconstruction at $z=0$ (top) and $z=1$ (bottom). Histograms show the distribution of the square root of the absolute difference between best-fit $\chi^2$ values with and without the BAO feature in a template for matter maps in real-space. Vertical dashed lines show the $3\sigma$ and $5\sigma$ detection significance.}
\label{fig:BAO_significance_matter_real}
\end{figure}

Another test we perform is to check which is the significance of the BAO detection in the case of different angular smoothing scales we are considering and to quantify the improvement after performing reconstruction. We do this by performing another fit to the measured matter monopole and quadrupole with a model that has no BAO feature. This model is constructed by taking $\Sigma_{\rm{nl}}\rightarrow \infty$.  Having obtained the $\chi^2_{\rm{NO\,BAO}}$ for each of our 500 simulations, we quantify the significance as the square root of $\Delta\chi^2=\chi^2_{\rm{NO\,BAO}}-\chi^2_{\rm{BAO}}$. In Figure \ref{fig:BAO_significance_matter_real} we show the histograms of $\sqrt{\Delta\chi^2}$ before (blue) and after (red) reconstruction for different map resolutions $\sigma$ at $z=0$ (top) and $z=1$ (bottom). 

We find that performing reconstruction greatly improves the significance of BAO detection. At $z=0$, using our matter maps in real-space covering a volume equal to $1\,h^{-1}\rm{Gpc}^3$, we find that 100\% (99\%) of our mocks shows better than $3\sigma$ ($5\sigma$) significance of detecting BAO after reconstruction. This result holds for all the considered map resolutions. We find similar results for reconstruction at $z=1$: 100\% (98.6\%) of our mocks shows a detection of the BAO with a significance above $3\sigma$ ($5\sigma$).

We also find that the improvement over the significance of BAO detection after reconstruction is greater for smaller values of the angular resolution, while the improvement decreases as we use larger values of angular resolution. Furthermore, the improvement is greater at $z=0$ than at $z=1$, as expected, since the non-linear effects that reconstruction partially removes are smaller at higher redshifts.

\subsubsection{Redshift-space}

By performing reconstruction over matter maps at $z=0$ in redshift-space we find that uncertainties in $\alpha$ decrease by $\sim 50\,\%$ after reconstruction, while the uncertainties in $\epsilon$ decrease by $\sim 40\,\%$ (see Table \ref{table:Table_matter}).  Recovered mean values of $\epsilon$ after reconstruction are at most 0.2\% away from the expected $\epsilon=0$ value and are within $5\sigma$ uncertainties considering the error on the mean, i.e. a total volume of 500 $(h^{-1}{\rm Gpc})^3$. On the other hand, we find a biased estimate of the recovered values in $\alpha$ that increases up to 0.5\% for larger smoothing scales $\sigma$. Even though this bias is statistically significant at a level of more than $5\sigma$ for a total simulated volume, it is still compatible with the expected value considering the typical volume of a future 21cm survey.

At $z=1$ we find that the uncertainties after reconstruction in $\alpha$ decrease by $\sim 30\,\%$, while for $\epsilon$ we find $\sim 50\,\%$ decrease. All the recovered values of both $\alpha$ and $\epsilon$ are consistent with the expected values. Small biases we find are within $5\sigma$ uncertainties for the full simulation volume.

As can be seen from Table \ref{table:Table_matter} and Figure \ref{fig:Alpha_epsilon_master} the errors on $\epsilon$ parameter are significantly smaller than the errors on $\alpha$. This result is in contrast to the results from real-space where the errors on $\alpha$ and $\epsilon$ are similar. The reason why this happens in redshift-space is due to the following fact. 
As described in Section \ref{sec:sims}, for matter maps in redshift-space we measure the monopole and the quadrupole along three different axes of our simulation. We then compute the covariance matrix by taking the average monopole and quadrupole along three different axes for each realisation. Since the scatter of quadrupole is high along three different axes in a particular realisation, taking the average reduces the variance and in effect makes the covariance matrix values smaller. Since the quadrupole is more sensitive to $\epsilon$, in turn this makes the uncertainties on $\epsilon$ smaller by roughly a factor of 3 compared to the case we use only one axis. On the other hand, since the monopole is using the spherically averaged information, the measured scatter between different axes is much smaller. In turn the scatter is not affected by averaging over three different axes. Being that the monopole is more sensitive to the $\alpha$ parameter, the constraints are very similar when considering only one axis or average over three axes. To summarise, if we use only one axis, which is a more realistic scenario and what we have done in real-space, we get the errors on $\epsilon$ to be comparable and larger than the errors on $\alpha$. In this case, the main reason why the constraints are similar for alpha and epsilon is the angular resolution, similar to the real-space consideration.

\begin{figure}[htbp]
\centering
\includegraphics[scale = 0.3]{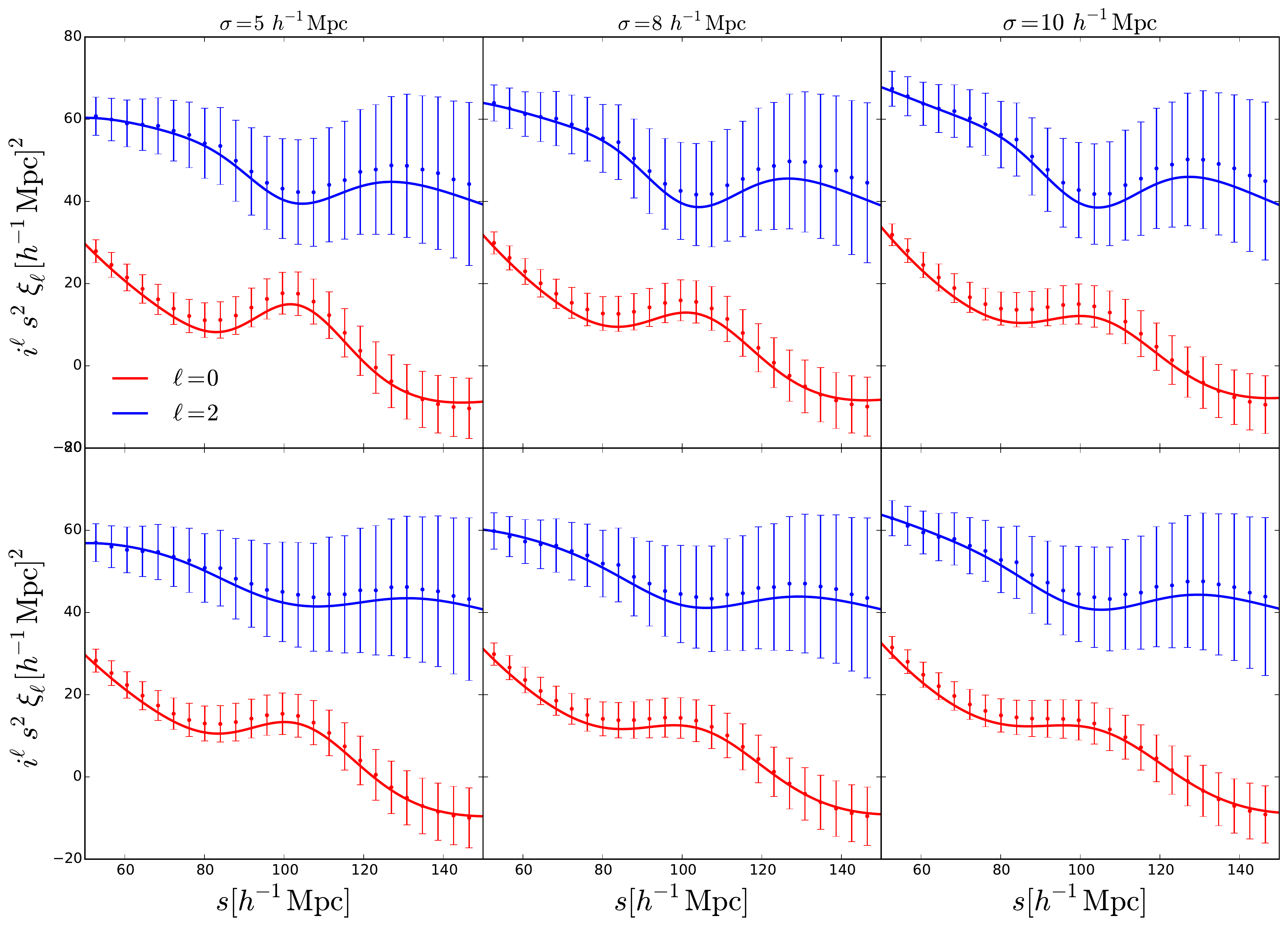}
\caption{The best-fit to monopole and quadrupole correlation function for matter maps in redshift-space at $z=0$ --  reconstructed (top) and unreconstructed (bottom).}
\label{fig:Best_fit_matter_redshift_z0}
\end{figure}

\begin{figure}[htbp]
\centering
\includegraphics[scale = 0.3]{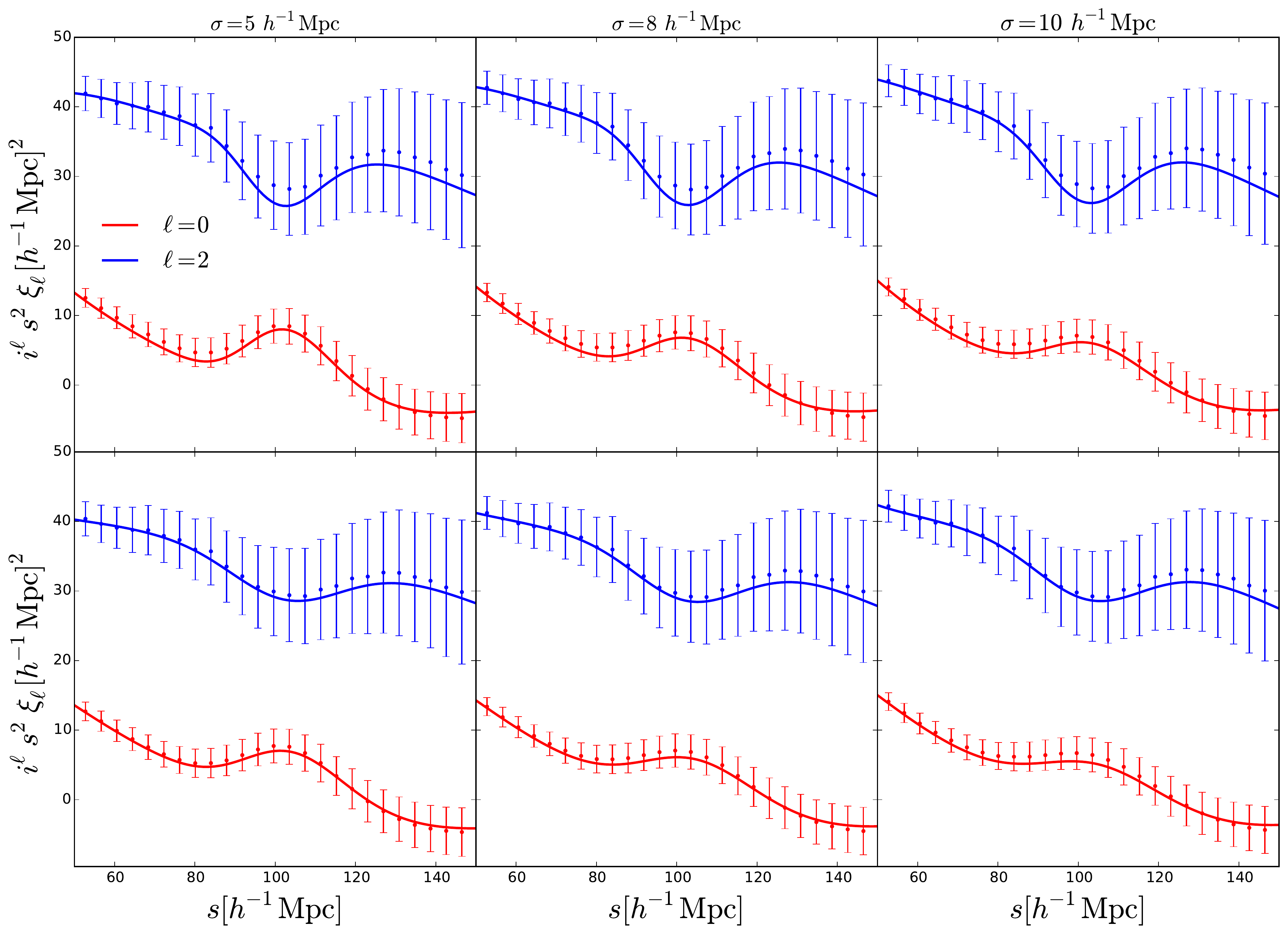}
\caption{The best-fit to monopole and quadrupole correlation function for matter maps in redshift-space at $z=1$ --  reconstructed (top) and unreconstructed (bottom).}
\label{fig:Best_fit_matter_redshift_z1}
\end{figure}

We show the best-fit model for matter maps in redshift space before and after reconstruction at $z=0$ in Figure \ref{fig:Best_fit_matter_redshift_z0} and at $z=1$ in Figure \ref{fig:Best_fit_matter_redshift_z1}. Similar to real-space, monopole is more broad for maps with larger angular smoothing scale, while in the quadrupole this effect is reversed. We find both monopole and quadrupole to be more pronounced after reconstruction. The effect is not so evident for the monopole, but we emphasize again that these results should be compared with the linear theory prediction (solid lines in Figure \ref{fig:Matter_real_z0_plots}, bottom panel). 

\subsection{Impact of angular resolution on measured distances}
\begin{figure}[htbp]
\centering
\includegraphics[scale = 0.6]{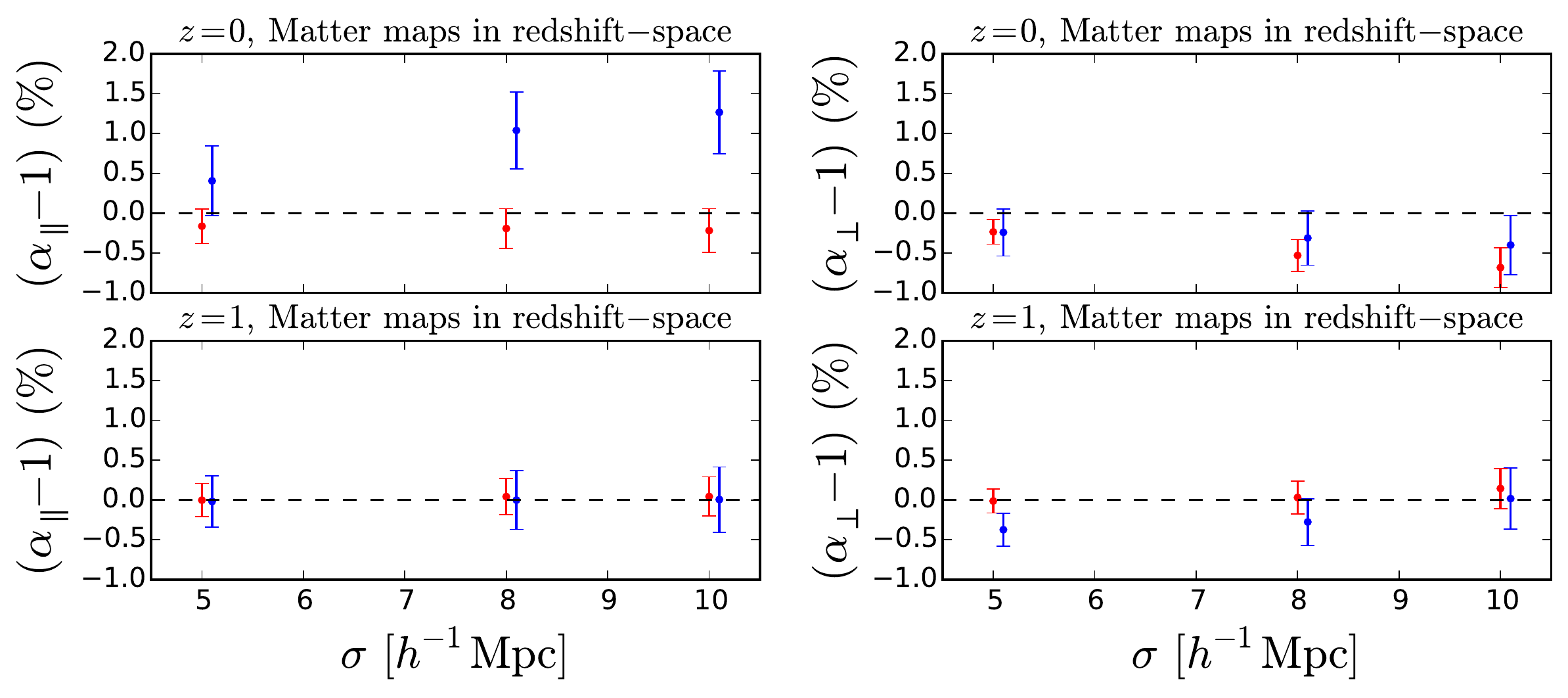}
\caption{Mean best-fit values of BAO dilation parameters $\alpha_{||}$ and $\alpha_{\perp}$ in redshift-space as a function of the smoothing scale $\sigma$. Blue points represent the non-reconstructed, while red points represent the reconstructed halo maps results. The error bars shown are the standard deviation from 500 simulations divided by $\sqrt{50}$ to show the uncertainty we would expect in a survey of $\sim50\,[h^{-1}\rm{Gpc}]^3$ volume size.}
\label{fig:alphas}
\end{figure}

Another useful parametrization of the position of the acustic scale is in terms of dilations along the line of sight and perpendicular to the line of sight,
\be
\alpha_{||} \equiv \frac{H_f r_{d,f}}{H r_d} \quad\text{and}\quad \alpha_{\perp}\equiv\frac{D_{A} r_{d,f}}{D_{A,f} r_{d}}\;\;,
 \ee 
which in real-space define $r^2 = \alpha_{||}^2 r_{||}^2 + \alpha_{\perp}^2 r_{\perp}^2$. A nice property of this parametrization is that it is linear in the cosmological parameters one wants to measure and therefore easier to interpret. The relation to the $\alpha$ and $\epsilon$ previously defined reads (see equation \ref{eq:alpha})
 \be
 \alpha = \alpha_{||}^{1/3} \alpha_{\perp}^{2/3}\quad \text{and}\quad \epsilon=\left(\frac{\alpha_{||}}{\alpha_{\perp}}\right)^{1/3}\;\;.
 \ee
As discussed in Section \ref{sec:analysis}, the effect of angular resolution is to further smooth the field perpendicularly to the line of sight, hence we expect the constraints on the angular diameter distance to be more affected by the value of $\sigma$ than the Hubble parameter (this has been extensively discussed in \cite{pacobao}).
This is shown in Figure \ref{fig:alphas} where we plot the constraints on $\alpha_{||}$ and $\alpha_{\perp}$ for dark matter in redshift space. 
While $\alpha_{||}$ benefit from reconstruction, both in terms of central value and $1\sigma$ error, almost independently of the additional angular smoothing, the same is not true for $\alpha_{\perp}$. At $z=0$, and for large values of $\sigma$, the best fit value of $\alpha_{\perp}$ is still biased with respect to the true value even after reconstruction. This indicates that non-linear shifts of the BAO are not well captured by the reconstruction procedure when too many modes are missing. We also note that the error on $\alpha_{\perp}$ is not reduced much by reconstruction when the angular resolution is too low. 
At $z=1$ the picture is somehow better, since change in the position of the acustic peak induced by gravity are less important as one moves to higher redshift. However the gain in errorbars after reconstruction is only marginal. 
This result actually questions how well the BAO could be measured in the transverse direction by a $21$ cm intensity mapping experiment (in \cite{pacobao} it was shown that SKA1-MID will not even detect the isotropic BAO peak at $z\ge1$). For instance, an experiment like CHIME has an angular resolution at $z=1$ comparable to our idealized single-dish case. 

\subsection{Halo maps}

We measure the BAO shift parameter $\alpha$ in each of our 500 halo maps before and after reconstruction using the template models outlined in Section \ref{sec:halos}. In Table \ref{table:Table_halos} we give the summary of the best-fit results for the isotropic BAO shift parameter $\alpha$ in real- and redshift-space at $z=0$. 

\begin{table}[ht!]
\centering
\begin{tabular}{cccc}

\hline \hline
\multicolumn{4}{c}{Halos -- Real Space}\\
\hline	

$\sigma[\hMpc]$	&	Reconstruction	&		$\alpha$				&	$\mean{\chi^2}/\mathrm{dof}$	\\ \hline
\multirow{2}{*}{5}	&	no	&	$	1.002	\pm	0.026	$	&	32.2/20	\\ 
	&	yes	&	$	1.000	\pm	0.016	$	&	27.0/20	\\ \cline{1-4}
\multirow{2}{*}{8}	&	no	&	$	1.004	\pm	0.028	$	&	31.5/20	\\
	&	yes	&	$	1.001	\pm	0.018	$	&	27.7/20	\\ \cline{1-4}
\multirow{2}{*}{10}	&	no	&	$	1.005	\pm	0.031	$	&	28.0/20	\\
	&	yes	&	$	1.001	\pm	0.020	$	&	26.2/20	\\ \hline

\multicolumn{4}{c}{Halos -- Redshift Space}\\ \hline

$\sigma[\hMpc]$	&	Reconstruction	&	$\alpha$	&	$\mean{\chi^2}/\mathrm{dof}$	\\ \hline
\multirow{2}{*}{5}	&	no	&	$	0.999	\pm	0.032	$	&	28.1/20	\\
	&	yes	&	$	0.998	\pm	0.019	$	&	26.3/20	\\ \cline{1-4}
\multirow{2}{*}{8}	&	no	&	$	0.998	\pm	0.034	$	&	27.6/20	\\
	&	yes	&	$	0.998	\pm	0.022	$	&	26.3/20	\\ \cline{1-4}
\multirow{2}{*}{10}	&	no	&	$	0.998	\pm	0.036	$	&	25.2/20	\\
	&	yes	&	$	0.999	\pm	0.024	$	&	24.5/20	\\ \hline \hline

\end{tabular}
\caption{Fitting results for halo maps in real- and redshift-space at $z=0$ for different map resolutions $\sigma$ (column 1). Column 3 show the mean and the standard deviation of BAO shift parameter $\alpha$. The mean $\chi^2/\mathrm{dof}$ is given in column 4. The errors shown for $\alpha$ are the standard deviations over 500 simulations. The actual errors on the means of the parameter $\alpha$ are the listed values divided by $\sqrt{500}$.}
\label{table:Table_halos}
\end{table}

\subsubsection{Real-space}
In Figure \ref{fig:Halos_alpha} we show the mean best-fit values of the shift parameter $\alpha$ as a function of the angular resolution $\sigma$ in real-space (left panel). We find that our reconstruction method works well in real-space and decreases the uncertainties on the parameter $\alpha$ by roughly 40\% compared to the unreconstructed case. The recovered shift parameter are consistent with the expected $\alpha=1$ and are within the uncertainties for all angular resolutions considered. 

The uncertainty on $\alpha$ shows an increases as we go to larger values of the angular resolution - $\sigma$. This is expected as we are here only considering the monopole in which the BAO peak gets less pronounced with larger angular resolution $\sigma$. We also expect that the constraints on $\alpha$ could get tighter and less $\sigma$-dependent if we were also able to use the information from the halo maps quadrupole. 

\begin{figure}[htbp]
\centering
\includegraphics[scale = 0.3]{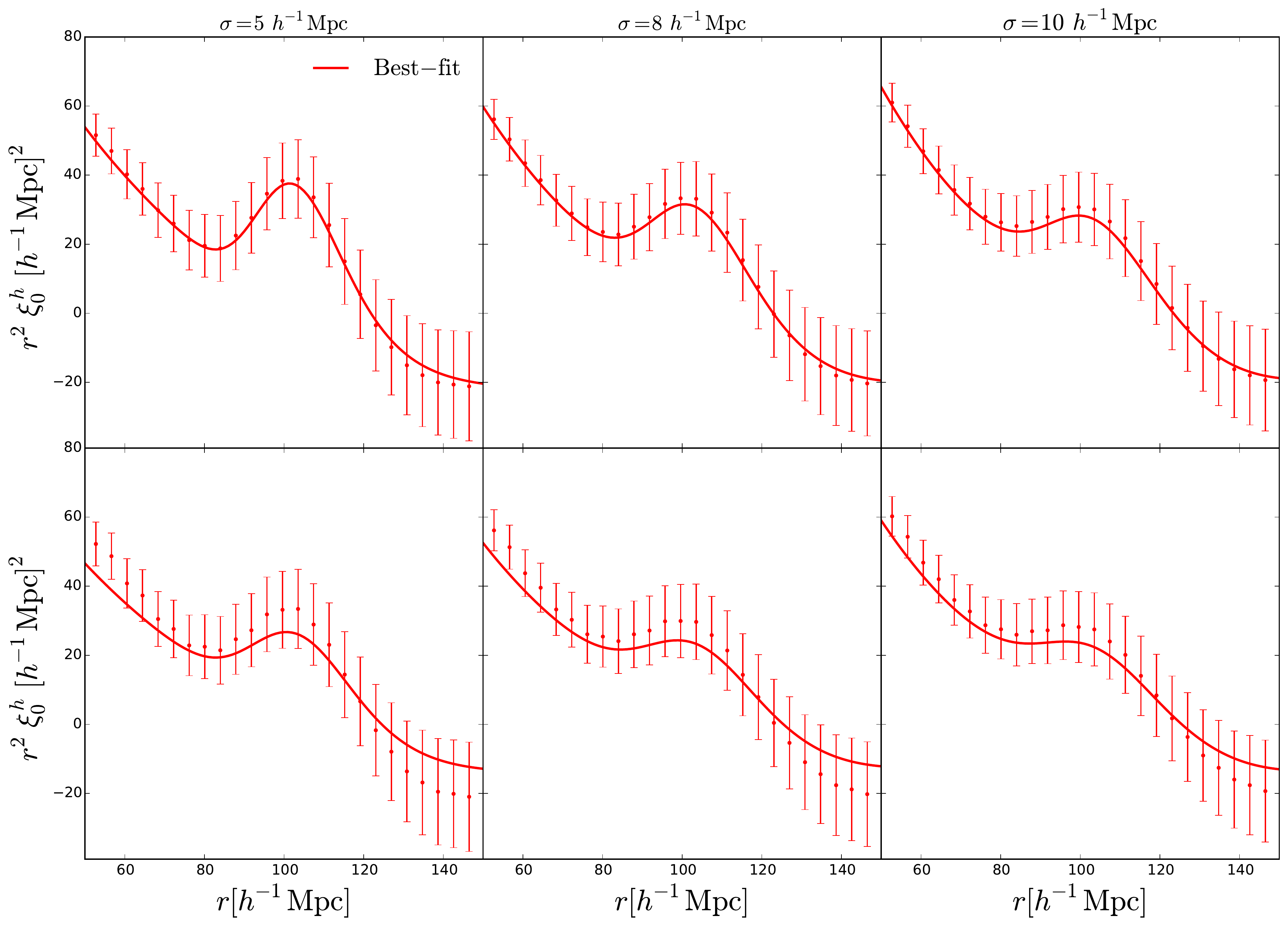}
\caption{The best-fit to monopole of the correlation function for halo maps maps in real-space at $z=0$ --  reconstructed (top) and unreconstructed (bottom).}
\label{fig:Best_fit_halos_real_z0}
\end{figure}

In Figure \ref{fig:Best_fit_halos_real_z0} we show the best-fit to the halo monopole in real-space. The BAO peak in the monopole gets more pronounced after reconstruction, suggesting our reconstruction method is able to partially remove the non-linear effects that cause the smearing of the BAO peak. With higher angular resolution $\sigma$ used, the monopole gets more broad, even in linear theory (as shown in upper panel in Figure \ref{fig:Halos_real_z0_plots}) and the effect of reconstruction is not so evident anymore.

\subsubsection{Redshift-space}
\begin{figure}[htbp]
\centering
\includegraphics[scale = 0.5]{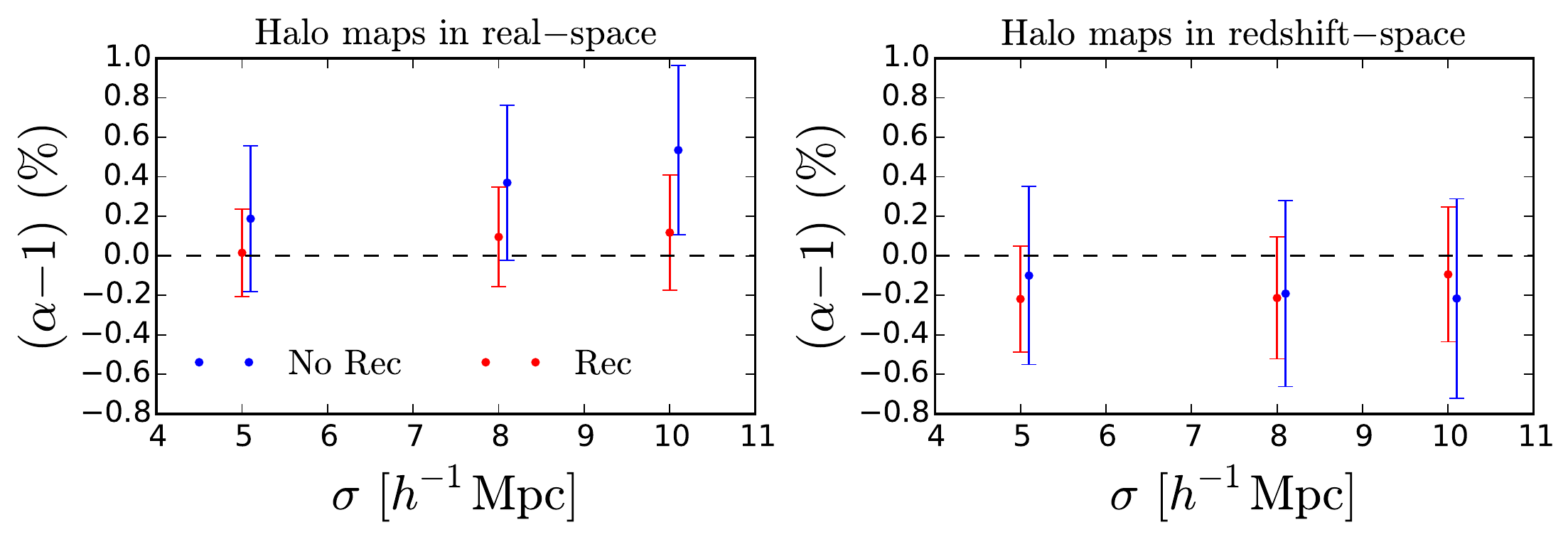}
\caption{Halo maps at $z=0$: Mean best-fit values of BAO shift parameter $\alpha$ in real- (left panel) and redshift-space (right panel) as a function of the angular resolution $\sigma$. Blue points represent the non-reconstructed, while red points represent the reconstructed halo maps results. The error bars shown are the standard deviation from 500 simulations divided by $\sqrt{50}$ to show the uncertainty we would expect in a survey of $\sim50\,[h^{-1}\rm{Gpc}]^3$ volume size.}
\label{fig:Halos_alpha}
\end{figure}

Figure \ref{fig:Halos_alpha} shows the mean best-fit values of shift parameter $\alpha$ as a function of angular resolution $\sigma$ in redshift-space (right panel). Similar to the results in real-space, we find the uncertainties on $\alpha$ after reconstruction decrease by 30\% -- for larger angular resolution, up to 40\% -- for smaller angular resolution $\sigma$. Our recovered mean values of $\alpha$ after reconstruction are within 0.2\% and consistent with the expected value 1. The biases we find are at the level of $3\sigma$ for the full 500 simulations volume.

\begin{figure}[htbp]
\centering
\includegraphics[scale = 0.3]{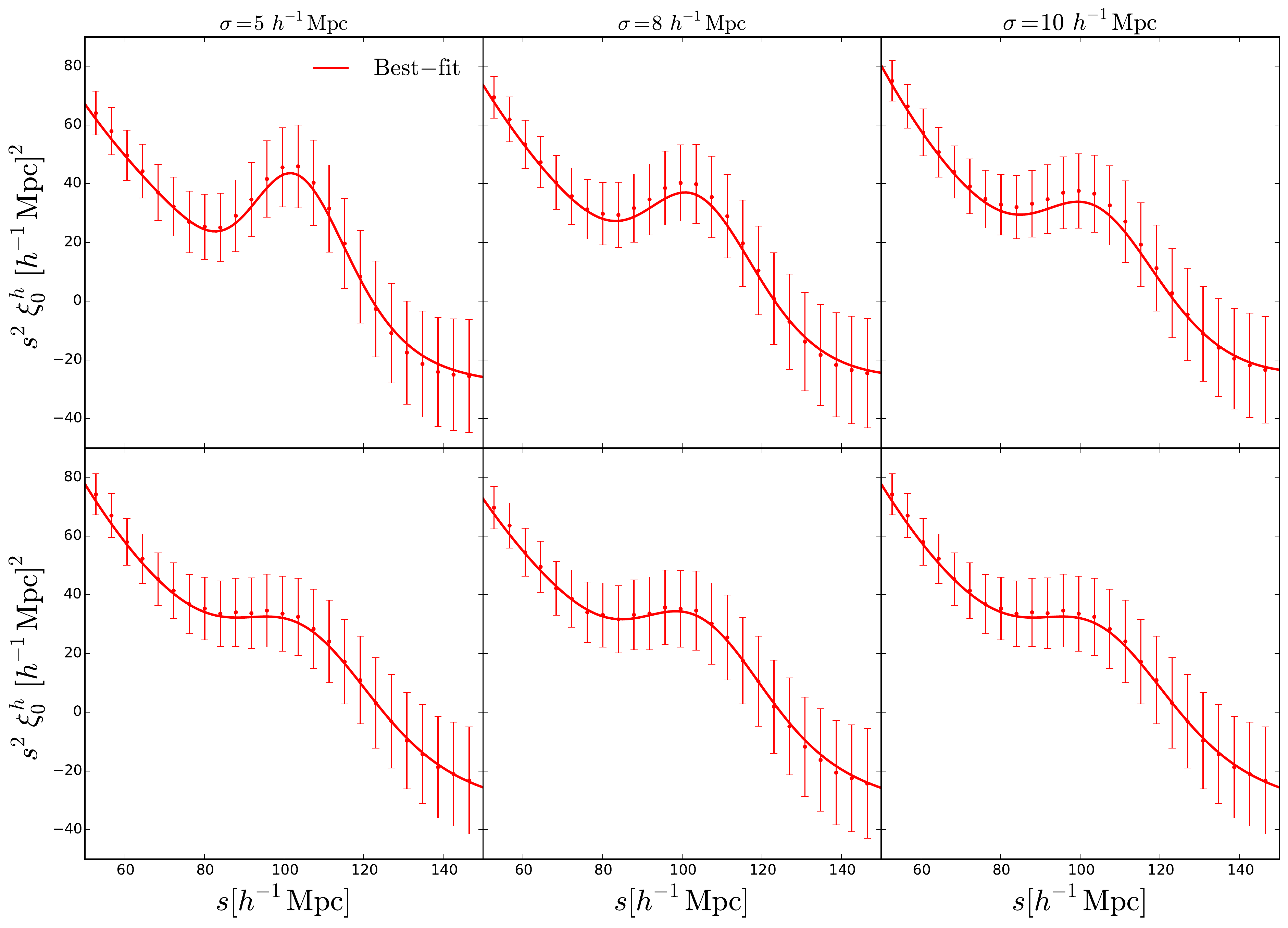}
\caption{The best-fit to monopole of the correlation function for halo maps in redshift-space at $z=0$ --  reconstructed (top) and unreconstructed (bottom).}
\label{fig:Best_fit_halos_redshift_z0}
\end{figure}

In Figure \ref{fig:Best_fit_halos_redshift_z0} we show the best-fit to halo correlation function in redshift space.  Similar to the results in real-space, the BAO peak in the monopole gets more pronounced after reconstruction, suggesting our reconstruction method is able to partially remove the non-linear effects that cause the smearing of the BAO peak in redshift-space too. Using higher angular resolution scales, the monopole gets less pronounce.

\section{Summary and Conclusions}
\label{sec:conclusions}

Perturbations in the early Universe produced sounds waves, called baryon acoustic oscillations, that propagated in the baryon-photon plasma until the recombination epoch. This phenomenon left its signature on the spatial distribution of matter and galaxies in the Universe as a peak (or set of wiggles) in the correlation function (power spectrum) that can be used as a standard ruler. The position of the BAO peak is very well constrained by CMB experiments, and by measuring it from low redshift cosmological probes (such as galaxy catalogues or 21cm intensity mapping observations) constraints on the value of the cosmological parameters, and therefore on the nature of dark energy, can be set. 

BAO are extremely robust cosmological probes with respect to systematic effects. Unfortunately, non-linear gravitational evolution tends to smear out that feature by inducing a damping and broadening on the BAO peak and produces a shift in its position. Those effects will induce a systematic bias on the derived value of the cosmological parameters (due to the peak shift) and will increase the error bars on those since the peak position will be less clear. \textit{Reconstruction} is a technique developed to undo (at least partially) the effects of non-linear gravitational evolution. 

The ultimate goal of standard reconstruction methods is to place galaxies in their initial positions. This is partially achieved by moving back galaxies by estimating the amplitude of the underlying density field and using the Zel'dovich approximation to compute displacement field \cite{Eisenstein_2007b}. This procedure has proven to be very successful and the constraints on the value of the cosmological parameters have improved after applying this technique \cite{Padmanabhan_2012, Anderson2014a}. 

However, there are cosmological observations that do not produce as output galaxy catalogues, but pixelated maps; an example of this kind of observations is a 21cm intensity mapping survey. While the power spectrum or correlation function inferred from these observations will be affected by non-linearities, in the same way galaxy surveys are, it is not obvious, a-priori, the way reconstruction should be performed on those density maps.

In this paper we have tested a new BAO reconstruction method that consists in moving pixels instead of galaxies. We work on a regular grid to compute the displacement field and then treat the grid cells as galaxies in the standard reconstruction. By doing this we avoid two interpolations of the displacement field -- one for the particles/galaxies and one for the uniform field of particles to compute the shifted field. Having the grid cells small enough, we recover the results from the standard method. 

The main features of this method are:
\begin{itemize}
\item It can be applied to both galaxy surveys and pixelated maps (e.g. 21cm intensity mapping observations).
\item In the limit of very small pixels it is equivalent to standard reconstruction method.
\item It is faster and easier to implement than the standard method.
\end{itemize}

We have tested this method against the standard one in the case of matter density field in real-space on a large set of large box-size numerical simulations. We varied the smoothing scale for the displacement field across a wide range and we find that the methods agree. We find no significant difference between the methods in the constraints on the position of the BAO peak that we recover.

We have then applied this method to the spatial distribution of matter and halos in both real- and redshift-space using the same numerical simulations. In all cases we take into account the pixelated nature of the observations by creating mock maps that we obtain by convolving the simulating field (matter or halos) with a 2-dimensional Gaussian beam that mimic the effect of the telescope primary beam. 

We use a theoretical template that parametrizes the effect of non-linearities on the broadening of the BAO peak, redshift-space distortions, FoG effect and basic instrumental effects such as the telescope beam size (embedded into the map resolution in our analysis) to fully model the measured correlation functions. 

Our findings can be summarized as follow:
\begin{itemize}
\item By reconstructing maps created from the spatial distribution of matter in real-space at $z=0$ and $z=1$ we find that the recovered values on $\alpha$ and $\epsilon$ are compatible with those expected, $\alpha=1$ and $\epsilon=0$, respectively. At $z=0$ the errors on $\alpha$ and $\epsilon$ decrease by $\sim50\%$ with respect to the case without reconstruction. We find the relative decrease in errors after reconstruction on both shift parameters vary within 5\% and shows no evident dependence on the map resolution. At $z=1$ the error on $\epsilon$ decreases by $\sim 30\,\%$, while the errors on $\alpha$ decrease by 30\% after reconstruction for smaller values of the smoothing scale, while it decreases by 15\% for larger values of the smoothing scale. 

\item By reconstructing maps created from the spatial distribution of matter in redshift-space at $z=0$ and $z=1$, we find that, at the level of expected precision of a future 21cm survey, the recovered values on $\alpha$ and $\epsilon$ are compatible with those expected, $\alpha=1$ and $\epsilon=0$. At $z=0$ the error on $\alpha$ error decreases by a 40-50\% with respect to the case without reconstruction, while the error of $\epsilon$ decreases by 35-45\% after reconstruction. We find the relative decrease in errors depends on the angular smoothing scale and the performance of reconstruction is more effective for smaller smoothing scales. 
At $z=1$ errors decrease by 30\% and 50\% for $\alpha$ and $\epsilon$, respectively, after reconstruction, and we find no significant dependence on the angular smoothing scale used. 

\item Using a different parametrisation of the BAO peak shifts ($\alpha_{\parallel},\alpha_{\perp}$), we see more clearly the effect of low angular resolution on the constraints on the BAO peak position along the line-of-sight and in the transverse direction. Our reconstruction of matter maps in redshift-space recovers the expected values of $\alpha_{\parallel}$ at the level of expected precision of a future 21cm survey and provides better constraints at both $z=0$ and $z=1$ with almost no dependence on the angular smoothing scale $\sigma$. In the case of $\alpha_{\perp}$, we find that with increasing angular smoothing scale, relative gains of reconstruction get smaller at $z=0$, while the situation is somewhat better at $z=1$.

\item By reconstructing maps created from the spatial distribution of halos in real-space at $z=0$ we find that the recovered value on $\alpha$ is compatible with 1. The error on the $\alpha$ after reconstruction decreases by a 40\% for the smallest smoothing scale, while it decreases by 30\% for the largest smoothing scale we used. We find the relative improvement on the constraints of $\alpha$ after reconstruction to depend on the smoothing scale.

\item By reconstructing maps created from the spatial distribution of halos in redshift-space at $z=0$ we find that the recovered value on $\alpha$ is compatible with 1. The error on $\alpha$ after reconstruction decreases by 40\% with respect to the case without reconstruction for the smallest, while the decrease is 30\% for the largest smoothing scale used.
\end{itemize}

In summary: in this paper we have tested a method in detail that is able to perform reconstruction in both galaxy surveys and pixelated maps as those from 21cm intensity mapping surveys. It consists in moving pixels rather than galaxies and it is equivalent to standard reconstruction in the limit of a very fine grid. We have tested this method by using a large set of numerical simulations and find an excellent agreement with theoretical expectations. We believe this method can be particularly useful to tighten the constraints on the value of the cosmological parameters from intensity mapping observations in the post-reionization era from surveys such as CHIME and SKA.

\acknowledgments
We thank Marko Simonovi\'c, Emiliano Sefusatti, David Alonso, Ariel Sanchez and Ravi Sheth for useful conversations. Numerical simulations have been carried out in the Ulysses cluster in SISSA (Trieste, Italy). The work of FVN is supported by the Simons Foundation. FVN and MV have been supported by the ERC Starting Grant ``cosmoIGM''. AO, FVN and MV are partially supported by INFN IS PD51 ``INDARK''. 


\bibliographystyle{JHEP}
\bibliography{References.bib}{}

\end{document}